\documentclass[12pt]{iopart}

\usepackage{iopams} 
\def\nbox#1#2{\vcenter{\hrule \hbox{\vrule height#2in
\kern#1in \vrule} \hrule}}
\def\sq{\,\raise.5pt\hbox{$\nbox{.09}{.09}$}\,} 
\def\sqb{\,\raise.5pt\hbox{$\overline{\nbox{.09}{.09}}$}\,}
 
\begin{document}

\title{Cosmological Dark Energy:\\
Prospects for a Dynamical Theory}

\author{Ignatios Antoniadis\footnote{On leave from CPHT (UMR CNRS 7644) 
Ecole Polytechnique, 91128 Palaiseau Cedex, France}}

\address{Department of Physics\\
CERN, Theory Division\\
CH-1211 Geneva 23, Switzerland}
\ead{ignatios.antoniadis@cern.ch}

\author{Pawel O. Mazur}

\address{Department of Physics and Astronomy,\\
University of South Carolina\\ Columbia SC 29208 USA}
\ead{mazur@mail.psc.sc.edu}

\author{Emil Mottola}

\address{T-8, Theoretical Division, MS B285\\
Los Alamos National Laboratory\\ Los Alamos, NM 87545 USA}
\ead{emil@lanl.gov}

\begin{abstract}
We present an approach to the problem of vacuum energy in cosmology,
based on dynamical screening of $\Lambda$ on the horizon scale.
We review first the physical basis of vacuum energy as a phenomenon 
connected with macroscopic boundary conditions, and the origin of
the idea of its screening by particle creation and vacuum polarization
effects. We discuss next the relevance of the quantum trace anomaly
to this issue. The trace anomaly implies additional terms in the low 
energy effective theory of gravity, which amounts to a non-trivial 
modification of the classical Einstein theory, fully consistent with 
the Equivalence Principle. We show that the new dynamical degrees of 
freedom the anomaly contains provide a natural mechanism for relaxing
$\Lambda$ to zero on cosmological scales. We consider possible signatures
of the restoration of conformal invariance predicted by the fluctuations of 
these new scalar degrees of freedom on the spectrum and statistics of 
the CMB, in light of the latest bounds from WMAP. Finally we assess the 
prospects for a new cosmological model in which the dark energy adjusts 
itself dynamically to the cosmological horizon boundary, and therefore
remains naturally of order $H^2$ at all times without fine tuning.

\end{abstract}

\centerline{Invited Contribution to \it{New Journal of Physics Focus 
Issue on Dark Energy}}
\vspace{.4cm}
\centerline{Preprint Nos. LA-UR-06-8631\ \ \ CERN-PH-TH/2006-251}

\maketitle

\section{Vacuum Fluctuations and the Cosmological Term}

Vacuum fluctuations are an essential feature of quantum theory. The attractive 
force between uncharged metallic conductors in close proximity, discovered 
and discussed by Casimir more than half a century ago, is due to the vacuum 
fluctuations of the electromagnetic field in the region between the conductors
\cite{Cas}. At first viewed perhaps as a theoretical curiosity, the Casimir 
effect is now being measured with increasing accuracy and sophistication in the 
laboratory \cite{Moh}. The Casimir force directly confirms the existence of 
quantum fluctuations, and the field theory methods for handling the ultraviolet 
divergences they generate, to obtain finite answers at macroscopic distance 
scales. When combined with the Equivalence Principle, also well established
experimentally, this success of relativistic quantum field theory should
permit the treatment of the effects of vacuum fluctuations at macroscopic 
distances in the context of general relativity as well.

In classical general relativity, the requirement that the field equations 
involve no more than two derivatives of the metric tensor allows for the 
possible addition of a constant term, the cosmological term $\Lambda$, 
to Einstein's equations,
\begin{equation}
R_a^{\ b} - {R\over 2} \,\delta_a^{\ b} + \Lambda \,\delta_a^{\ b} 
= \frac{8\pi G}{c^4}\, T_a^{\ b}.
\label{Ein}
\end{equation}
If transposed to the right side of this relation, the $\Lambda$ term corresponds 
to a constant energy density $\rho_{\Lambda} = c^4\Lambda /8\pi G$ and isotropic 
pressure $p_{\Lambda} = - c^4\Lambda/8\pi G$ permeating all of space uniformly, 
and independently of any localized matter sources. Hence, even if the
matter $T_a^{\ b} = 0$, a cosmological term causes spacetime to become 
curved with a radius of curvature of order $\vert\Lambda\vert^{-{1\over 2}}$.

In purely classical physics there is no natural scale for $\Lambda$. 
Indeed if $\hbar =0$ and $\Lambda = 0$, there is no fixed length scale 
at all in the vacuum Einstein equations, $G/c^4$ being simply a conversion
factor between the units of energy and those of length. Hence $\Lambda$ 
may take on any value whatsoever with no difficulty (and with no explanation) 
in classical general relativity. 

As soon as we allow $\hbar \neq 0$, there is a quantity with the 
dimensions of length that can be formed from 
$\hbar, G$, and $c$, namely the Planck length,
\begin{equation}
L_{pl} \equiv \left( \frac{\hbar G}{c^3}\right)^{\frac{1}{2}}
= 1.616 \times 10^{-33}\,{\rm cm}.
\label{Lpl}
\end{equation}
Hence when quantum theory is considered in a general
relativistic setting, the quantity
\begin{equation}
\lambda \equiv \Lambda L_{pl}^2 = \frac{\hbar G \Lambda}{c^3}
\label{Lnum}
\end{equation}
becomes a dimensionless pure number, whose value one might expect 
a theory of gravity incorporating quantum effects to address.

Some eighty years ago W. Pauli was apparently the first to consider the
question of the effects of quantum vacuum fluctuations on the the curvature 
of space \cite{Pau,Ner}. Pauli recognized that the sum of zero point energies of 
the two transverse electromagnetic field modes {\it in vacuo},
\begin{equation}
\rho_{\Lambda} = 2 \int^{L_{min}^{-1}}\frac{d^3 {\bf k}}{(2\pi)^3} 
\frac{\hbar \omega_{\bf k}}{2}
 = \frac{1}{8\pi^2} \frac{\hbar c}{L_{min}^{\ \ 4}} = -p_{\Lambda}
\label{zeropt}
\end{equation}
contribute to the stress-energy tensor of Einstein's theory as would an 
effective cosmological term $\Lambda > 0$. Since the integral (\ref{zeropt}) 
is quartically divergent, an ultraviolet cutoff $L_{min}^{-1}$ of (\ref{zeropt}) 
at large $\bf k$ is needed. Taking this short distance cutoff $L_{min}$ to be of 
the order of the classical electron radius $e^2/mc^2$, Pauli concluded that if 
his estimate were correct, Einstein's theory with this large a $\Lambda$ would 
lead to a universe so curved that its total size ``could not even reach to 
the moon." If instead of the classical electron radius, the apparently natural 
but much shorter length scale of $L_{min} \sim L_{pl}$ is used to cut off the 
frequency sum in (\ref{zeropt}), then the estimate for the cosmological term in 
Einstein's equations becomes vastly larger, and the entire universe would be 
limited in size to the microscopic scale of $L_{pl}$ (\ref{Lpl}) itself, in even 
more striking disagreement with observation.

Clearly Pauli's estimate of the contribution of short distance modes of
the electromagnetic field to the curvature of space, by using (\ref{zeropt}) 
as a source for Einstein's eqs. (\ref{Ein}) is wrong. The question is why. 
Here the Casimir effect may have something to teach us. The vacuum zero point 
fluctuations being considered in (\ref{zeropt}) are the same ones that contribute 
to the Casimir effect, but this estimate of the scale of vacuum zero point energy, 
quartically dependent on a short distance cutoff $L_{min}$, is certainly {\it not} 
relevant for the effect observed in the laboratory. In calculations of the Casimir 
force between conductors, one subtracts the zero point energy of the electromagnetic 
field in an infinitely extended vacuum (with the conductors absent) from the 
modified zero point energies in the presence of the conductors. It is this 
{\it subtracted} zero point energy of the electromagnetic vacuum, depending upon 
the {\it boundary conditions} imposed by the conducting surfaces, which leads 
to experimentally well verified results for the force between the conductors. In 
this renormalization procedure the ultraviolet cutoff $L_{min}^{-1}$ drops out, 
and the distance scale of quantum fluctuations that determine the magnitude of the 
Casimir effect is not the microscopic classical electron radius, as in Pauli's original 
estimate, nor much less the even more microscopic Planck length $L_{pl}$, but 
rather the relatively {\it macroscopic} distance $d$ between the conducting 
boundary surfaces. The resulting subtracted energy density of the vacuum between
the conductors is
\begin{equation}
\rho_v = -\frac{\pi^2}{720}\, \frac{\hbar c}{d^4} \,.
\label{Casimir}
\end{equation} 
This energy density is of the opposite sign as (\ref{zeropt}), leading 
to an attractive force per unit area between the plates of 
$0.013$ dyne/cm$^2$ $(\mu m/d)^4$, a value which is both independent of the 
ultraviolet cutoff $L_{min}^{-1}$, and the microscopic details of the atomic 
constituents of the conductors. This is a clear indication, confirmed 
by experiment, that the {\it measurable} effects associated with 
vacuum fluctuations are {\it infrared} phenomena, dependent upon 
macroscopic boundary conditions, which have little or nothing to do 
with the extreme ultraviolet modes or cutoff of the integral in (\ref{zeropt}).

By the Equivalence Principle, local short distance behavior in a mildly
curved spacetime is essentially equivalent to that in flat spacetime.
Hence on physical grounds we should not expect the ultraviolet 
cutoff dependence of (\ref{zeropt}) to affect the universe in the large
any more than it affects the force between metallic conductors in the 
laboratory. 

In the case of the Casimir effect a constant zero point energy of the
vacuum, no matter how large, does not affect the force between the plates.
In the case of cosmology it is usually taken for granted that any effects of
boundary conditions can be neglected.  It is not obvious then what should 
play the role of the conducting plates in determining the magnitude of 
$\rho_v$ in the universe, and the magnitude of any effect of quantum zero 
point energy on the curvature of space has remained unclear from Pauli's 
original estimate down to the present. In recent years this has evolved from 
a question of fundamental importance in theoretical physics to a central one 
of observational cosmology as well. Observations of type Ia supernovae at 
moderately large redshifts ($z\sim 0.5$ to $1$) have led to the conclusion 
that the Hubble expansion of the universe is {\it accelerating} \cite{SNI}. 
According to Einstein's equations this acceleration is possible if and only 
if the energy density and pressure of the dominant component of the universe 
satisfies the inequality, 
\begin{equation}
\rho + 3 p \equiv \rho\  (1 + 3 w) < 0\,.
\label{accond}
\end{equation}
A vacuum energy with $\rho > 0$ and $w\equiv p_v/\rho_v = -1$ leads to an 
accelerated expansion, a kind of ``repulsive" gravity in which the relativistic
effects of a negative pressure can overcome a positive energy density in
(\ref{accond}). Taken at face value, the observations imply that some $74\%$ of 
the energy in the universe is of this hitherto undetected $w=-1$ dark 
variety. This leads to a non-zero inferred cosmological term in Einstein's 
equations of 
\begin{equation}
\Lambda_{\rm meas} \simeq (0.74)\, {3 H_0^2\over c^2} 
\simeq 1.4 \times 10^{-56}\ {\rm cm}^{-2}
\simeq  3.6 \times 10^{-122}\ {c^3 \over \hbar G}\,.
\label{cosmeas}
\end{equation}
Here $H_0$ is the present value of the Hubble parameter, approximately 
$73\, {\rm km/sec/Mpc} \simeq 2.4 \times 10^{-18}\, {\rm sec}^{-1}$. The last 
number in (\ref{cosmeas}) expresses the value of the cosmological dark 
energy inferred from the SN Ia data in terms of Planck units, 
$L_{\rm pl}^{-2} = {c^3 \over \hbar G}$, {\it i.e.} the dimensionless
number in (\ref{Lnum}) has the value 
\begin{equation}
\lambda \simeq 3.6 \times 10^{-122}\,.
\label{lmeas}
\end{equation} 
Explaining the value of this smallest number in all of physics is the
basic form of the ``cosmological constant problem."

We would like to emphasize that the naturalness problem posed by the very
small value of the cosmological vacuum energy $\lambda$ of (\ref{cosmeas})
arises only when quantum fluctuations ($\hbar \neq 0$) and gravitational 
effects ($G \neq 0$) are considered {\it together}. As we have already noted, 
if the universe were purely classical, $L_{pl}$ would vanish and $\Lambda$, 
like the overall size or total age of the universe, could take on any 
value whatsoever without any technical problem of naturalness. Likewise as 
the Casimir effect makes clear, if $G=0$ and there are also no boundary effects 
to be concerned with, then the cutoff dependent zero point energy of 
flat space (\ref{zeropt}) can simply be subtracted, with no observable 
consequences. A naturalness problem arises only when quantum vacuum fluctuations 
are weighed by gravity, and the effects of vacuum zero point energy on the large 
scale curvature of spacetime are investigated. This is a problem of the quantum 
aspects of gravity at {\it macroscopic} distance scales, very much greater than 
$L_{pl}$. Indeed,  what is measured in the supernova data (\ref{cosmeas}) is {\it not} 
directly the energy density of the vacuum (\ref{zeropt}) at all, but rather the 
geometry of the universe at very large distance scales. The dark energy content 
of the universe and the equation of state $p_v/\rho_v \approx - 1$ are inferred 
from the observations, by assuming the validity of (\ref{accond}) and therefore
of Einstein's equations (\ref{Ein}) as the effective theory of gravity 
applicable at cosmological distances.

The treatment of quantum effects at distances much larger than any ultraviolet 
cutoff is precisely the context in which effective field theory (EFT) techniques 
should be applicable. The EFT point of view is the one we shall adopt for this 
article. This means that we assume that we do not need to know every detail of
physics at extremely short distance scales of $10^{-33}$ cm or even $10^{-13}$ cm
in order to discuss cosmology at $10^{28}$ cm scales. What is important instead
is the Equivalence Principle, {\it i.e.} invariance under general coordinate 
transformations, which greatly restricts the form of any EFT of gravity. 
In his search for field equations for a metric theory with universal matter 
couplings, which incorporate the Equivalence Principle automatically but which 
is no higher than second order in derivatives of the metric, Einstein was 
using what we would now recognize as EFT reasoning. In an EFT treatment 
quantum effects and any ultraviolet (UV) divergences they generate at very 
short distance scales are absorbed into a few, finite low energy effective 
parameters, such as $G$ and $\Lambda$. 

General coordinate invariance of the low energy EFT does requires a more 
careful renormalization procedure than a simple normal ordering subtraction, 
which suffices for the original Casimir calculations in flat space. The UV 
divergent terms from the stress tensor must be isolated and carefully 
removed in a way consistent with the Equivalence Principle to extract physical 
effects correctly. These more general renormalization procedures, involving 
{\it e.g.} proper time, covariant point splitting or dimensional regularization 
have been developed in the context of quantum field theory in curved spacetime
 \cite{BiD}. The non-renormalizability of the classical Einstein theory poses 
no particular obstacle for an EFT approach. It requires only that certain 
additional terms be added to the effective action to take account of UV 
divergences which are not of the form of a renormalization of $G$ or $\Lambda$. 
The result of the renormalization program for quantum fields and their vacuum 
energy in curved space is that general relativity can be viewed as a low energy 
quantum EFT of gravity, provided that the classical Einstein-Hilbert classical 
action is augmented by these additional terms, a necessary by product of which 
is the quantum trace anomaly of massless fields \cite{anom}. We do not review 
the technology of renormalization of the stress tensor here, referring the
interested reader to the literature for details \cite{BiD}. We shall make 
extensive use of one important result of the renormalization of the stress 
tensor however, namely the trace anomaly and its effects at large distance 
scales. Hence it is the renormalization of the quantum stress tensor in 
curved space which provides the rigorous basis for an EFT approach to 
gravity at distance scales much larger than $L_{Pl}$.

The essential physical assumption in an EFT approach is the hypothesis
of {\it decoupling}, namely that low energy physics is independent of very 
short distance degrees of freedom and the details of their interactions. All 
of the effects of these short distance degrees of freedom is presumed to be
encapsulated in a few phenomenological coefficients of the infrared relevant 
terms of the EFT. Notice that this will not be the case in gravity if the 
low energy $\Lambda$ relevant for dark energy and cosmology depends upon the 
quantum zero point energies of all fields up to some UV cutoff, as in 
(\ref{zeropt}). If (\ref{zeropt}) is to be believed, then the introduction of 
every new field above even some very large mass scale would each contribute its 
own zero point energy of the order of $L_{min}^{-4}$ and generate additional 
terms relevant to the large scale curvature of spacetime. Clearly this  
contradicts any intuitive notion of decoupling of very massive states from low 
energy physics. Despite the severe violation of decoupling this represents, 
the usual presumption is that the ``natural" scale for $\Lambda$ 
is of order unity in Planck units, {\it i.e.} $\lambda \sim 1$.

In order to make the naturalness problem of small numbers and large
hierarchies more precise, 't Hooft gave the intuitive notion of 
fine tuning a technical definition \cite{tHo}. According to his formulation, 
a parameter of an EFT can be small naturally, only if setting it equal to zero 
results in a larger symmetry of the theory. Then quantum corrections will not 
upset the hierarchy of scales, once imposed. An example of such a naturally 
small parameter is the ratio of the pion mass to the $\rho$ meson or nucleon 
mass in QCD. In the limit of vanishing $u$ and $d$ quark masses QCD possesses 
an enhanced $SU_{ch}(2)$ chiral symmetry, and $m_{\pi} \rightarrow 0$ in the 
chiral limit. Even if the quark masses are finite, this is a ``soft" breaking 
of $SU_{ch}(2)$, and Goldstone's theorem protects $m_{\pi}$ from receiving large 
loop corrections at the otherwise natural scale of the strong interactions, 
$m_{\rho}\simeq 770$ MeV. Of course, this approximate symmetry does not enable 
one to predict the magnitude of chiral symmetry breaking in the strong interactions, 
and the actual small values of $m_{\pi} \simeq 140$ MeV or the $u$ and $d$ quark 
masses in QCD remain unexplained. However, the enhanced symmetry as these masses 
go to zero does permit the soft breaking scale of chiral symmetry to be quite 
different in principle from the UV cutoff scale of the pion EFT, since at 
least any large hierarchy of scales and a small value of $m_{\pi}/m_{\rho}$ 
is not automatically upset by quantum loop corrections.

In the case of the cosmological term $\Lambda$, the problem is that Einstein's 
theory does not possess any apparent enhanced symmetry if $\Lambda$ is set equal
to zero. This is hardly surprising since as we have already pointed out,
the classical theory contains no natural scale with which to compare $\Lambda$, 
any value being equally allowed {\it a priori}. Supersymmetry does not help 
here since in order to account for the absence of supersymmetric partners 
to the standard model particles at low energies, supersymmetry must be 
spontaneously broken at an energy scale no lower than approximately $1$ TeV. 
Then the natural scale of $\Lambda$ is still some $57$ orders of magnitude 
larger than that measured by the acceleration of the universe in (\ref{cosmeas}). 
Similar considerations apply to any new symmetry invoked at very short distances, 
and the problem persists even in apparently more microscopic description of quantum 
effects in gravity at the Planck scale. This impasse emphasizes once again that the 
problem of vacuum energy arises on macroscopic distance scales, and suggests that 
there is some basic ingredient missing in our EFT estimate of supposedly very weak 
quantum effects in gravity at {\it large} distances. 

A fine tuning problem potentially related to the naturalness problem
for the dimensionless $\lambda$ posed by (\ref{cosmeas}) is that of the
dimensionless ratio of $\rho_{\Lambda}$ to the closure density,  
\begin{equation}
\Omega_{\Lambda} \equiv  \frac{8\pi G\rho_{\Lambda}}{3c^2 H_0^2} =
\frac {c^2 \Lambda_{\rm meas}}{3H_0^2} \simeq 0.74 \,.
\label{OLam}
\end{equation}
According to the conventional adiabatic expansion history of the universe,
in which little or no entropy is generated during large epochs of time,
$\Omega_{\Lambda}$ is strongly time dependent, being very much smaller than 
unity at early times when matter or radiation dominates, but approaching 
unity exponentially rapidly at late times, as these other components are 
diluted by the expansion. Thus the value inferred from observations 
(\ref{OLam}) would seem to imply that we are living in a very special epoch 
in the history of the universe, when the vacuum energy has grown to be an
appreciable fraction  ($74\%$) of the total, but just before the matter 
and radiation have been redshifted away completely, as conventional 
adiabatic theory indicates they soon will be, exponentially rapidly. This
``cosmic coincidence problem," independently of the naturalness problem 
of the very small value of $\lambda$ also suggests that something 
basic may be missing from current cosmological models. The first indication
as to what that element may be emerges from consideration of quantum effects 
in curved spacetimes such as de Sitter spacetime.

\section{Quantum Effects in de Sitter Spacetime}

The simplest example of accelerated expansion is a universe composed 
purely of vacuum energy, {\it i.e.} $\Omega_{\Lambda} = 1$. 
This is de Sitter spacetime with $H^2 = \Lambda c^2/3$ a constant.
In classical general relativity de Sitter spacetime is the stable 
maximally symmetric solution to Einstein's equations (\ref{Ein}) 
with positive $\Lambda$ and $T_a^{\ b} = 0$ otherwise. In spatially 
flat, homogeneous and isotropic Robertson-Walker (RW) form de Sitter
spacetime has the line element,
\begin{equation}
ds^2 = - c^2 d\tau^2 + a^2(\tau)\, d{\bf x}^2\,.
\label{RWflat}
\end{equation}
Here $\tau$ is the proper time of a freely falling observer and
$a(\tau)$ is the RW scale factor, determined from
the Friedman equation, 
\begin{equation}
H^2 \equiv \Bigl( \frac{\dot a}{a}\Bigr)^2 
= \frac{8 \pi G}{3c^2}\ \rho\,.
\label{rw}
\end{equation}
with the overdot denoting differentiation with respect to $\tau$.
The cosmological constant is included in the right hand side of (\ref{rw})
as a vacuum energy contribution with equation of state, $\rho_{\Lambda}= 
- p_{\Lambda} = c^4\Lambda /8\pi G$. With $\Omega_{\Lambda} = 1$,
and $H^2 = c^2 \Lambda/3$ a constant, the RW scale factor is
\begin{equation}
a_{de S} (\tau) = e^{H\tau}\,,
\label{RWdS}
\end{equation}
in de Sitter spacetime.

At the beginning of modern cosmology the de Sitter model was proposed 
to account for the Hubble expansion \cite{DES}. This model was soon abandoned 
in favor of Friedman-LeMa\^{i}tre-Robertson-Walker (FLRW) models, with apparently 
more physical ordinary matter and radiation replacing the unknown $\rho_{\Lambda}$ 
as the dominant energy density in the universe. Unlike the eternal expansion of 
(\ref{RWdS}), these FLRW models possess an origin of time at which the universe 
originates at a spacelike singularity of infinite density, {\it i.e.} a big bang. 
Evidence for the radiation relic of this primordial explosion was discovered in 
the cosmic microwave background radiation (CMB). The CMB is remarkably 
uniform (to a few parts in $10^5$) over the whole sky, and hence its 
discovery immediately raised the question of how this uniformity could 
have been established. In the classical FLRW models with their spacelike 
initial singularity, there is no possibility of causal contact between 
different regions of the present microwave sky, and hence these models
possess a causality or horizon problem. In the 1960's Sakharov and Gliner 
observed that a de Sitter epoch in the early universe could remove this causality 
problem of the initial singularity in the standard matter or radiation dominated 
cosmologies \cite{Sak}. In the 1980's such a de Sitter phase in the early 
history of the universe was proposed under the name of inflation, with a vacuum 
energy scale $\hbar H$ of the order of $10^{15}$ GeV, associated with the 
unification scale of the strong and electroweak interactions \cite{Guth}. A key 
success of inflationary models is that the small deviations from exact homogeneity 
and isotropy of the CMB, now measured in beautiful detail by the COBE and WMAP 
satellites \cite{WMAP}, can originate from quantum zero point fluctuations in 
the de Sitter phase, with a magnitude determined by the ratio $\hbar H/M_{Pl}c^2$ 
\cite{LidL}. Thus if inflationary models are correct, microscopic quantum physics 
at the tiny unification scale of $10^{-29}$ cm is responsible for the large 
scale classical inhomogeneities of the matter distribution in the universe 
at $10^{25}$ cm and above. The supposition of inflationary models that  
microscopic quantum physics might be responsible for the classical
distribution of galaxies at cosmological distance scales is no less 
remarkable now than when it was first proposed a quarter century ago.

Despite the central importance of quantum fluctuations in the de Sitter
phase of inflationary models, and the cosmological vacuum energy
responsible for the inflationary epoch itself rooted in quantum 
theory, current inflationary models remain quite classical 
in character. For example one large class of models involves the 
``slow roll" of a postulated (and so far unobserved) scalar inflaton 
field in a classical potential $V$ \cite{LidL}. In models of this kind the 
classical evolution of the inflaton must be slow enough to allow for a 
sufficiently long-lived de Sitter phase, to agree with the presently observed 
size and approximate flatness of the universe, a condition that requires
fine tuning of parameters in the model. The value of the cosmological 
$\Lambda$ term responsible for the present acceleration of the universe 
(generally assumed to be identically zero prior to the supernova data) is 
unexplained by scalar inflaton models and requires additional fine tuning. 
Indeed fine tuning issues are characteristic of all phenomenological 
treatments of the cosmological term, with a large hierarchy of scales and no 
fundamental quantum theory of vacuum energy.

Beginning in the 1980's, quantum fluctuations and their backreaction 
effects in de Sitter spacetime were considered in more detail by a 
number of authors, including the present ones \cite{deSp,deSt,Fluc,deSMM}. 
Several of these studies indicated that fluctuations at the horizon 
scale $c/H$ could be responsible for important effects on the classical 
de Sitter expansion itself. Based on these studies, a mechanism for relaxing 
the effective value of the vacuum energy to zero over time dynamically was 
proposed \cite{deSp,time}. Although a fully satisfactory cosmological model 
based on these ideas does not yet exist, a dynamical theory of vacuum energy 
still appears to be the most viable alternative to fine tuning or
purely anthropic considerations for the very small but non-zero value 
of $\lambda$. For a recent overview of approaches we refer the reader
to ref. \cite{Nob}.

Given the incomplete and scattered nature of the results in the technical 
literature, the need for a review of the subject accessible to a wider 
audience has been apparent for some time. With the discovery of dark energy 
in the universe, and the recent twenty-fifth anniversary of inflationary 
models, we have thought it worthwhile to review at this time the status and 
prospects for a dynamical resolution to the problem of vacuum energy. In this 
section, we begin by reviewing the infrared quantum effects in de Sitter space, 
which first suggested a dynamical relaxation mechanism, in roughly the historical 
order that they were first discussed.

\subsection{Particle Creation in de Sitter Space}

A space or time dependent electric field creates particles.
J. Schwinger first studied this effect in QED in a series of classic 
papers \cite{Schw}. Later, Parker, Fulling, Zel'dovich and many
others realized that a time dependent gravitational metric 
should also produce particles \cite{BiD}. The study of these effects
formed the beginning of the subject of quantum fields in curved 
spacetime. From this point of view the exponential de Sitter
expansion (\ref{RWdS}) provides a time dependent background
field which can create particle pairs from the ``vacuum,"
converting the energy of the classical gravitational background
into that of particle modes. 

Both the concept of ``vacuum" and ``particle" in a background gravitational
(or electromagnetic) field merit some comment. Since particle number 
generally does not commute with the Hamiltonian in spacetime dependent 
backgrounds, a unique definition of a ``vacuum" state devoid of 
particles does not exist in this situation. In flat space with no background 
fields, relativistic wave equations have both positive energy (particle) 
solutions and negative energy (anti-particle) solutions, which are clearly 
distinguishable by time reversal symmetry. In time dependent backgrounds 
this is no longer the case and the two solutions mix on the scale 
of the time variation. Physically this is because in quantum theory a 
particle cannot be localized to a region smaller than its de Broglie 
wavelength. When this wavelength becomes large enough to be of the same 
order of the scale of spacetime variation of the background field, the 
particle concept begins to lose its meaning and it is better to think of 
matter as waves rather than as particles. Of course, this is precisely 
the wavelike effect of quantum matter which we characterize as 
``particle" creation.  A fair amount of the technical literature on 
particle creation is concerned with defining what a particle is and what 
sort of detector in a particular state of motion can detect it. Much of 
this is an irrelevant to the backreaction problem, and will not concern 
us here. Provided one sticks to questions of evaluating conserved currents 
in well-defined initial states, one can bypass any technical discussion of 
definitions of ``particles" {\it per se}. The solutions of the wave equation of 
the quantum field(s) undergoing the particle creation, and the evolution 
of its electric current or energy-momentum tensor once renormalized
are well-defined, and independent of arbitrary definitions of particle number.
For a consistent physical definition of adiabatic particle number see ref. \cite{AMP}.

Let us consider the electromagnetic case in more detail.
To motivate the discussion, we note that there is an analogy to 
the vacuum energy problem in electromagnetism as well, the ``cosmological 
electric field problem" \cite{time}. It consists of the elementary observation 
that Maxwell's equations {\it in vacuo, i.e.} in the absence of all
sources admit a solution with constant, uniform electric field 
${\bf E}_{cosm}$ of arbitrary magnitude and direction. Why then are 
all electric fields we observe in nature always associated with localized 
electrically charged matter sources? Why do we not observe some macroscopic 
uniform electric field pointing in an arbitrary direction in space? 

A non-zero ${\bf E}_{cosm}$ selects a preferred direction in space. 
Nevertheless, and perhaps somewhat surprisingly, relativistic particle 
motion in a uniform constant electric field has precisely the {\it same} 
number of symmetry generators (ten) as those of the usual zero field 
vacuum. These $10$ generators define a set of modified spacetime
symmetry transformations, leaving ${\bf E}_{cosm}$ fixed, whose algebra 
is isomorphic to that of the Poincar\'{e} group \cite{elec}. In this 
respect field theory in a spacetime with a constant, non-dynamical 
${\bf E}_{cosm} \neq 0$ is similar to field theory in de Sitter spacetime 
with a constant, non-dynamical $\Lambda \neq 0$. The isometry group of 
de Sitter spacetime is $O(4,1)$, which also has $10$ generators, exactly 
the same number of flat Minkowski spacetime. In the absence of a unique
choice of vacuum in a spacetime dependent background, the point of view
often adopted is to choose the state with the largest possible symmetry
group permitted by the background. In de Sitter spacetime this maximally
symmetric state is called the Bunch-Davies (BD) state \cite{BD}.

Ordinarily one would say that the ground state should also be the
one of lowest energy, and as a non-zero electric field has
a non-zero energy density, the ground state should be that with
${\bf E}_{cosm} = 0$. However, once we appeal to an energy
argument we must admit that we do not know the absolute energy
of the vacuum, and because of (\ref{zeropt}) must allow for an
arbitrary shift of that zero point, perhaps compensating for
the electromagnetic field energy. Again $\Lambda = 0$ is 
similar to $ {\bf E}_{cosm} = 0$ in having zero energy in flat
space. As with $\Lambda$ classically we are free to set ${\bf E}_{cosm} = 0$ 
by an appropriate boundary condition at very large distances, but 
this leaves unanswered the question of why this is the appropriate 
boundary condition for our universe. The suggestion that $\Lambda$
may be regarded also as a constant of integration has been made
by a number of authors \cite{intg}. In quantum theory the boundary
condition which determines a constant of integration in the
low energy description requires information about the macroscopic 
quantum state of the system at large distance scales. Is there any 
evidence that quantum effects are relevant to the question of the 
cosmological electric field or of $\Lambda$?

The answer is almost certainly yes, because of quantum vacuum polarization
and particle creation effects. Suppose that ${\bf E}_{cosm}$ were not zero. 
Then the quantum vacuum of any charged matter fields interacting with it 
is polarized. This is described by a polarization tensor,
\begin{equation}
\Pi^{ab} (x, x'; {\bf E}) = i \bigl\langle 
{\cal T} j^a(x) j^b(x') \bigr\rangle_{\bf E}, 
\label{polj}
\end{equation}
where $j^a$ is the charge current operator and the expectation value is evaluated 
in a state with classical background field $\bf E$. This could be taken to be
the state of maximal symmetry allowed by the background, in particular a state
with discrete time reversal symmetry. The time ordering symbol ${\cal T}$ enforces 
the Feynman boundary conditions on the polarization operator. These boundary 
conditions are {\it not} time symmetric. Time-ordering (with an $i \epsilon$
prescription in the propagator) defines a different polarization tensor or
Feynman Green's function from anti-time-ordering (with a $-i\epsilon$ prescription). 
This time asymmetry is built into quantum theory by the demands of causality which 
distinguishes retarded from advanced effects in the polarization function (\ref{polj}).
Correspondingly, the polarization operator $\Pi^{ab}$ contains two pieces, an even 
and an odd piece under time reversal. The even piece describes the polarizability 
of the vacuum, since the vacuum fluctuations of virtual charge pairs may be thought 
of as giving rise to an effective polarizability of the vacuum (dielectric constant). 
The time reversal odd piece describes the creation of {\it real} particle anti-particle 
pairs from the vacuum by the Schwinger mechanism. The creation of real charged pairs 
means that a real current flows ${\bf j} \ge 0$ in the direction of the electric field, 
even if none existed initially. This implies that the electric field does work at a rate, 
${\bf j}\cdot{\bf E}$. To the extent that this power cannot be recovered because the 
created particles interact and lose the coherence they may have had in their initial 
state, this is the rate of energy {\it dissipation}. The vacuum can behave then very
much like a normal conductor with a finite conductivity and resistivity, 
due to the random, uncorrelated motions of its fluctuating charge carriers. 

At the same time, the electric field is diminished by the Maxwell 
equation,
\begin{equation}
{\partial {\bf E} \over \partial t} = -  \langle{\bf j}\rangle 
\label{Max}
\end{equation}
in the case of exact spatial homogeneity of the average current. 
The important question is that of the time scale of the effective dissipation. 
Does the degradation of the coherent electric field take place rapidly 
enough to effectively explain why there is no observed ${\bf E}_{cosm}$ today? 
The original Schwinger calculation of the decay rate of the vacuum into charged 
pairs involves a tunneling factor, $\exp(-\pi m^2c^3/ eE\hbar)$ for the creation 
of the first pair from the vacuum under the assumption of particular initial
conditions. An exponential tunneling factor like this would greatly suppress the 
effect. However, if any charged matter is present initially, {\it i.e.} if we
are not in precisely the maximally symmetric ``vacuum" state, the charges are 
accelerated, radiate, and pair produce without any tunneling suppression factor. 
Thus the time scale for an {\it induced} cascade of particle pairs to develop 
and degrade the electric field energy in any state save a carefully tuned initial 
vacuum state will be very much faster than the Schwinger spontaneous vacuum rate.
Hence on physical grounds there is an {\it instability} of the vacuum in a 
background electric field and extreme sensitivity to boundary conditions.
These are the necessary conditions for the spontaneous breaking of time 
reversal symmetry \cite{time}. 

The particle creation effects in homogeneous electric fields, including the 
backreaction of the mean current on the field through (\ref{Max}) have been 
studied in some detail in the large $N$ or mean field approximation \cite{CKM}. 
In this approximation the direct scattering between the created particles is 
neglected. Inclusion of scattering processes opens additional channels and 
faster time scales for dissipation, as in classical plasmas or Boltzmann gases. 
Allowing for these dissipative processes should permit any long range coherent 
${\bf E}_{cosm}$ present initially to relax to very small values on time scales 
much shorter than a Hubble expansion time, $H_0^{-1}$.

The electromagnetic analogy is that of a giant capacitor discharging. 
Any cosmological electric field initially present in the universe eventually 
shorts itself out, and degrades to zero, when the vacuum polarization effects 
described by $\Pi^{ab}$ and realistic particle interactions are taken into account. 
The actual value of the electric field at late times may then be very much less 
than  its ``natural" value of $m^2c^3/\hbar e$ or any other scale related to short
distance physics. The real and imaginary parts of $\Pi^{ab}$ are related by a 
dispersion relation which is one form of a fluctuation-dissipation 
theorem for the electrically polarized quantum vacuum. It is simply a consequence 
of causality, and the existence of a positive Hilbert space of quantum states 
that quantum vacuum fluctuations and quantum vacuum dissipation are inseparably 
related. One necessarily implies the other. This is the modern, relativistically
covariant formulation for quantum fields of the relation between an equilibrium 
quantity (mean square displacement) and a time asymmetric dissipative quantity
(diffusion coefficient or viscosity) first discussed by Einstein in his theory 
of Brownian movement \cite{brown}.

These observations are of a very general nature, and apply equally well to vacuum 
fluctuations in a gravitational background field. Since the geometry couples to 
the energy-momentum stress tensor, it is the fluctuations in this quantity which 
govern the dynamics of the gravitational field, and we are led to consider the 
corresponding polarization tensor,
\begin{equation}
\Pi^{abcd} (x, x'; {\bar g}) = 
i \bigl\langle {\cal T} T^{ab}(x) T^{cd}(x') \bigr\rangle_{\bar g}, 
\label{polt}
\end{equation}
where ${\bar g}_{ab}(x)$ represents some classical background metric, for
example that of de Sitter spacetime (\ref{RWflat}-\ref{RWdS}). This polarization 
tensor may be handled by exactly the same techniques as (\ref{polj}), and the 
analogous fluctuation-dissipation theorem relating its real and imaginary parts 
may be proven \cite{Fluc}. If the background ${\bar g}_{ab}(x)$ possesses a 
timelike Killing field, and therefore a Euclidean continuation with periodicity 
$\beta = \hbar/k_{_B}T$, it is natural to introduce the Fourier transform 
with respect to the corresponding static coordinate time difference $t-t'$. 
Defining the Fourier transforms of the symmetric and anti-symmetric parts 
of (\ref{polt}) by
\begin{equation}
\eqalign{
\int_{-\infty}^{\infty} dt\ \bigl\langle \bigl\{ T^{ab}(x), T^{cd}(x')\bigr\}_+ 
\bigr\rangle e^{i \omega (t-t')} &= S^{abcd}({\bf r}, {\bf r'}; \omega),\cr
\int_{-\infty}^{\infty} dt\ \bigl\langle \bigl[ T^{ab}(x), T^{cd}(x')\bigr]_- 
\bigr\rangle e^{i \omega (t-t')} &= D^{abcd}({\bf r}, {\bf r'}; \omega),\cr}
\label{symant}
\end{equation}
we find that the two pieces are related via:
\begin{equation}
D^{abcd}({\bf r}, {\bf r'}; \omega) = \tanh \left({\beta \omega \over 2}\right)\  
S^{abcd}({\bf r}, {\bf r'}; \omega).
\label{fluc}
\end{equation}
One can show that the condition for particle creation effects to occur in 
the background electric or gravitational field is just the condition that 
the time asymmetric piece of the exact polarization function {\it diverges} for 
large $t-t'$, in particular that:
\begin{equation}
D^{abcd}({\bf r}, {\bf r'}; \omega) \propto \omega^{-1} ,\qquad \omega 
\rightarrow 0.
\label{sing}
\end{equation}
In fact it is precisely the residue of this $\omega^{-1}$ pole which determines 
the particle creation rate in the adiabatic limit of slowly varying backgrounds
\cite{Fluc}. This {\it singular} behavior at low frequencies means that the 
background is unstable to small perturbations, and is the signal for
spontaneous breaking of time reversal symmetry. This sensitivity to infrared
fluctuations in $\Pi^{abcd}$ is why the inclusion of quantum fluctuations and 
correlation functions higher than the average $\langle T_a^{\ b}\rangle$
can change the behavior of a macroscopic quantum system over long times.

In cosmology the Friedman equation (\ref{rw}) together with the equation 
of covariant energy conservation,
\begin{equation}
{\dot \rho} + 3\, H\, (\rho + p) = 0, 
\label{ener}
\end{equation}
imply that:
\begin{equation}
{\dot H} = - \frac{4 \pi G}{c^2} \ (\rho + p). 
\label{hd}
\end{equation}
This relation from Einstein's equations is to be compared to the Maxwell 
eq. (\ref{Max}). In both cases there is a classical static background 
that solves the equation trivially, namely $H$ or $\bf E$ a constant 
with zero source terms on the right hand side. In the case of (\ref{hd}) 
this is de Sitter spacetime with $\rho_{\Lambda} + p_{\Lambda} = 0$. In order 
to exhibit explicitly the static nature of de Sitter spacetime, we make the 
coordinate transformation,
\numparts
\begin{eqnarray}
2H\tau &=& 2H t + \ln\left(1-H^2r^2/c^2\right)\\
\vert{\bf x}\vert &=& r\,e^{-H\tau} = r\,e^{-Ht}\,
\left(1-H^2r^2/c^2\right)^{-\frac{1}{2}}\,,
\label{ctrans}
\end{eqnarray}
\endnumparts
to bring the de Sitter line element (\ref{RWflat}) with (\ref{RWdS})
into the form,
\begin{equation}
ds^2\Big\vert_{de S} = -(c^2 - H^2r^2)\, dt^2 + 
\frac{dr^2}{1-H^2r^2/c^2} + r^2 (d\theta^2 + \sin^2\theta\, d\phi^2)\,,
\label{dSstatic}
\end{equation}
in which the metric is independent of the static time variable $t$. 
The translation $\partial_t$ defines a isometry of de Sitter
spacetime or Killing field, which is timelike for $r < c/H$.
This is the time translation invariance of de Sitter space which 
permits the introduction of a time stationary state with an
equilibrium thermal distribution at the inverse temperature $\beta$
of (\ref{fluc}). In the representation (\ref{dSstatic}) 
the static nature of de Sitter space and the existence
of the observer horizon at $r_{_H} = c/H$ are manifest, but an arbitrary 
particular point in space $r=0$ is chosen as the origin, so that 
the spatial homogeneity of the Robertson-Walker coordinates (\ref{RWdS}) 
is no longer manifest. Spacetime events for $r > r_{_H}$ are causally
disconnected and unobservable to a freely falling observer at
the origin of static coordinates. 

In the static coordinates (\ref{dSstatic}) the de Sitter metric becomes space 
dependent rather than time dependent, just as a constant electric field can be 
expressed either in a time dependent (${\bf A} = - {\bf E} t$) or static 
($A_0 = - {\bf E} \cdot {\bf r}$) gauge. The effects of the de Sitter background 
on the polarization of the vacuum and particle creation should be independent of 
the coordinates or gauge. In each case the static $H$ or $\bf E$ background field 
is stable to classical matter perturbations. Classical matter (if charged in the 
$\bf E$ case) is simply accelerated, and swept out by the constant electric or 
gravitational field. In the de Sitter case classical matter (obeying $\rho + p > 0$) 
is redshifted away by (\ref{ener}). However if quantum matter fluctuations possess a 
spectrum with a singular $\omega^{-1}$ dependence for small frequencies (\ref{sing}), 
then the classically stable background with an event horizon is unstable to quantum 
fluctuations. In that case the system will be driven away from the quasi-static 
initial state towards a final state in which the classical field energy has been 
dissipated into matter or radiation field modes. In ref. \cite{Fluc} it was argued 
that the polarization function corresponding to scalar ({\it i.e.} metric trace) 
perturbations of the de Sitter background has precisely the required singular 
$\omega^{-1}$ behavior. This behavior describes the response of the system 
to perturbations on length scales of order of the horizon size $r_H$ or larger. 
Then de Sitter spacetime satisfies the condition for dissipation of curvature 
stress-energy into matter and radiation modes, much as the electric field
background considered previously. 

Actually computing the evolution away from the initial state by this effect 
requires that we go beyond the simple replacement of the quantum stress tensor 
$T_a^{\ b}$ by its expectation value $\langle T_a^{\ b}\rangle$ and consider 
fluctuations about the mean as in (\ref{polt}), as well as possibly higher 
correlation functions of the stress tensor as well. This is not an easy task. 
Nevertheless it is worth noting that whereas the Schwinger suppression factor 
for vacuum tunnelling occurs in electrodynamics because there are no massless 
charged particles, massless particles do couple gravitationally, and would be 
expected to dominate the dissipative process. In both cases the creation of 
matter particle pairs with ${\bf j\cdot E} > 0$ and $\rho + p > 0$ causes the 
background field parameter, $E_{cosm}$ or $H$ to decrease. In fact, unlike the 
electric current which is a vector and may change sign, $\rho + p$ for realistic 
matter or radiation is always positive, so we should expect $H$ to decrease 
monotonically to zero, without the plasma oscillations that can occur in the 
electrodynamic case \cite{CKM}. 

The problems with implementing these promising ideas in a realistic model are
mainly technical. As we have already noted, fluctuations about the mean
stress tensor and their backreaction on the mean geometry must be taken into
account in a consistent calculation. Infrared divergences are encountered in any 
direct attempt to evaluate the diagrams contributing to the dissipation process 
perturbatively. Infrared problems of this kind due to long range forces are well 
known in the low frequency hydrodynamic limit of plasmas and gauge field theories 
at finite temperatures \cite{LeB}. In many analogous situations of this kind, the 
$\omega \rightarrow 0$ limit is outside the range of validity of perturbative 
expansions, due to collective effects. In hydrodynamics the long time behavior 
of correlation functions (\ref{fluc}) are non-perturbative, requiring at the 
least a resummation of perturbative processes. The problem of dissipating
vacuum energy by microscopic particle creation effects may be
compared in order of difficulty with the problem of evaluating the 
viscosity of water and the damping of eddy currents in a stream
from the electronic structure and interactions of the H$_2$O molecule.

Although a full calculation including self-interactions and backreaction 
along these lines has not been done, even in electromagnetism, the formal 
similarity between charge carrier fluctuations and Maxwell's equation for the 
displacement current on the one hand, and fluctuations in stress-energy and 
Einstein's equations (\ref{Ein}) on the other suggests that a dissipative 
relaxation of the vacuum energy into ordinary matter and radiation is possible 
via this mechanism. The bulk viscosity of the cosmological ``fluid" of
vacuum fluctuations is the quantity controlling this dissipation. Quantum 
fluctuations on or near the horizon scale are the relevant ones which need 
to be handled in a consistent, reliable, non-perturbative framework, in order 
to convert the coherent vacuum energy of de Sitter space to matter/radiation 
modes on the time scales relevant for cosmology.

\subsection{Thermodynamic Instability of de Sitter Spacetime}

A second set of considerations points to the role of dynamical quantum 
effects on the horizon scale in de Sitter space. The existence of the observer 
horizon at $r=r_H$ leads in conventional treatments to a Hawking temperature for 
freely falling observers \cite {GH}. The Hawking temperature is closely related 
to the particle creation effect since both depend upon a mixing between positive 
and negative frequency components of quantum fields in de Sitter space at the 
horizon scale. It is worth emphasizing that like the Casimir effect, this is a 
global effect of the spacetime, where the horizon now sets the scale, playing 
the role of the boundary conditions on the conducting plates in (\ref{Casimir}).
In the BD state particle creation and annihilation effects are exactly 
balanced in a precisely time symmetric manner and a configuration 
formally similar to that of thermodynamic equilibrium is possible. The Hawking 
temperature in de Sitter space,
\begin{equation}
T_{_H} = \frac{\hbar H}{2\pi k_{_B} } = \frac{\hbar c}{2\pi k_{_B}}\,
\sqrt{\frac{\Lambda}{3}}
\label{deSHaw}
\end{equation}
is the temperature a freely falling detector would detect in the
BD state. This temperature becomes the one which enters the
fluctuation-dissipation formulae (\ref{symant})-(\ref{fluc})
of the previous subsection in de Sitter space. Although this 
temperature is very small for $\lambda \ll 1$, a thermodynamic 
argument similar to Hawking's original one for black holes implies 
that the BD equilibrium state in de Sitter space is thermodynamically 
{\it unstable} \cite{deSt}.

Consider as the ``vacuum" state in de Sitter spacetime that state
defined by the analytic continuation of all of its Green's functions
by $t \rightarrow it$. Since the resulting geometry in (\ref{dSstatic})
is a space of uniform positive curvature, {\it i.e} a sphere $S^4$
with $O(5)$ isometry group and radius $cH^{-1}$, the Euclidean Green's 
functions are periodic in imaginary time with period $2\pi/H$. Continuing 
back to Lorentzian time, the BD state defined by this analytic 
continuation is a thermal state with temperature $T_{_H}$ \cite{deSGP}, 
invariant under the full $O(4,1)$ de Sitter symmetry group \cite{BiD}. Because 
of that symmetry, the expectation value of the energy-momentum tensor of any matter 
fields in this state must itself be of the form $\rho = - p =$ constant. This 
is the BD state usually assumed, tacitly or explicitly by inflationary model 
builders. 

The existence of such a maximally symmetric state does not guarantee its 
stability against small fluctuations, any more than the existence of a 
state invariant under the $10$ isometries of space with a constant electric 
field guarantee the stability of the vacuum with an electric field. In fact, 
both the energy within one horizon volume, and the entropy of the de Sitter 
horizon $S_H$ are {\it decreasing} functions of the temperature, {\it i.e.},
\numparts
\begin{eqnarray}
E_{_H} &= \rho_{\Lambda} V_H = \frac{c^5}{2GH} 
= \frac{\hbar c^5}{4\pi Gk_{_B}T_H}\,;\\
S_{_H} &= k_{_B} \frac{A_H}{4 L_{pl}^2}  =
\frac{\pi c^5 k_{_B}}{\hbar GH^2} = 
\frac{\hbar c^5}{4\pi G k_{_B} T_H^2} =
\frac{3\pi}{\lambda}\, k_{_B}\,.
\label{ESdS}
\end{eqnarray}
\endnumparts
Hence, by considering a small fluctuation in the Hawking temperature of the 
horizon which (like the black hole case) causes a small net heat exchange 
between the region interior to the horizon and its surroundings, we find that 
this interior region behaves like a system with negative heat capacity
\cite{deSt},
\begin{equation}
\frac{dE_{_H}}{dT_{_H}} = - \frac {E_{_H}}{T_{_H}} =
 -\frac{3\pi\,k_{_B}}{\lambda}  < 0\,.
\label{heat}
\end{equation}
However, negative heat capacity is impossible for a stable system in 
thermodynamic equilibrium. It corresponds to a runaway process in which any
infinitesimal heat exchange between the regions interior and exterior to
the horizon will drive the system further away from its equilibrium
configuration. Since the choice of origin of static coordinates in
de Sitter space is arbitrary, the entire space is unstable to
quantum/thermal fluctuations in its Hawking temperature nucleating a
kind of vacuum bubble at an arbitrary point, breaking the global 
$O(4,1)$ de Sitter invariance down to $O(3)$ rotational invariance. 

This thermodynamic consideration is consistent with the previous one based 
on particle creation and the fluctuation-dissipation theorem, and again
suggests that collective quantum effects on the horizon scale are the
relevant ones. The enormously negative heat capacity for de Sitter space 
in (\ref{heat}) for $\lambda \ll 1$, if taken literally suggests that the 
time scale for the instability to develop may not be exponentially large given
any initial perturbation. Despite this signal of thermodynamic instability, 
an evaluation of $\Pi^{abcd}$ and full dynamical analysis of the fluctuations 
about the BD state in de Sitter space in real time has not been 
carried out, again mostly for technical reasons. The framework for such a 
linear response analysis has been laid down only recently in ref. \cite{val}. 
A linearized analysis exhibiting an unstable mode would still be only the 
first step in a more comprehensive non-perturbative treatment of its effects.

\subsection{Graviton Fluctuations in de Sitter Spacetime}

A third route for investigating quantum effects in de Sitter spacetime
leading to the same qualitative conclusions is through studies of the 
fluctuations of the metric degrees of freedom themselves. The propagator 
function describing these metric fluctuations is a function of two spacetime 
points $x^a$ and $x^{\prime\,a}$. In flat spacetime, assuming a Poincar\'e 
invariant vacuum state, the propagator becomes a function only of the 
invariant distance squared between the points, {\it i.e.} $(x-x')^2$. 
Likewise full invariance of the graviton propagator under the global 
$O(4,1)$ de Sitter isometry group is a necessary condition for the 
gravitational vacuum in de Sitter spacetime to exist and to be stable 
to perturbations. Using such covariant methods to evaluate the propagator 
encounters a problem however. If one requires de Sitter invariance by 
computing the Euclidean propagator on $S^4$ with $O(5)$ invariance group,
and then analytically continuing to de Sitter spacetime, then one 
obtains a graviton propagator with rather pathological properties. 
Both the transverse-tracefree (spin-2) and trace (spin-0) 
projections of the Feynman propagator grow without bound both at large 
spacelike and large timelike separations \cite{AIT, AMJMP, AlTur}. Since 
this propagator leads to infrared divergences in physical scattering
processes \cite{AMJMP}, this large distance behavior of the propagator
function cannot be removed by a gauge transformation. This infrared
behavior is a striking violation of cluster decomposition properties 
of the de Sitter invariant vacuum state.

A similar situation had been encountered before in de Sitter spacetime,
in the quantization of a massless, minimally coupled free scalar field,
obeying the wave equation, $\sq \Phi = 0$. A covariant construction 
of the propagator function for $\Phi$ meets the obstacle that $\sq$ 
has a normalizable zero mode (namely a constant mode) on the Euclidean 
continuation of de Sitter space, $S^4$. Hence a de Sitter invariant 
propagator inverse for the wave operator $\sq$ does not exist \cite{AlFol}. 
Formally projecting out the problematic mode leads to a propagator
function which grows logarithmically for large spacelike or timelike
separations of the points $x$ and $x'$. Since the wave equation for 
spin-2 graviton fluctuations is identical to that of two massless, 
minimally coupled scalar fields (one for each of the two polarization 
states of the graviton) in a certain gauge \cite{AIT, Ford}, it 
is not surprising that it shares many of the same features as the 
massless scalar case. 

If one abandons the method of Euclidean continuation from $S^4$ and 
quantizes either the massless scalar or the graviton by canonical 
methods, starting on a fixed spacelike surface rather than imposing 
global de Sitter invariance, one finds a Feynman propagator function 
of $x$ and $x'$ that is {\it not} de Sitter invariant \cite{Hig}. In other 
words, canonical quantization of either the massless scalar or graviton 
field necessarily breaks de Sitter invariance, and no de Sitter invariant 
vacuum exists in either case. 

The closest analog of this behavior in flat spacetime is that of massless 
scalar field in two dimensions. The Feynman propagator $G(x,x')$ for a 
massless field in $1+1$ dimensions satisfies
\begin{equation}
-\sq G(x, x') = \hbar\, \delta^2(x-x')\,.
\label{flatwo}
\end{equation}
If we require Lorentz invariance, the propagator must be a function of
the invariant $s = (x-x')^2$. In that case, the wave operator
$\sq$ becomes an ordinary differential operator in $s$ and the only
solutions to the homogeneous wave equation (for $x\neq x'$) are
ln $s$ and a constant. The coefficient of the ln $s$ solution is 
fixed by the delta function in (\ref{flatwo}), and we find
\begin{equation}
G(x,x') = -\frac{\hbar}{4\pi} \ln [\mu^2 (x-x')^2]
\label{logtwo}
\end{equation}
with $\mu$ an arbitrary constant which acts as an infrared cutoff. 
The logarithmic growth at large distances and infrared cutoff
dependence of the propagator implies that free asymptotic 
particle states do not exist for a massless field in two dimensions. 
In a canonical treatment the origin of the infrared problem is the 
constant $k=0$ Fourier mode of the field, which grows linearly with 
time. A Lorentz invariant normalizable ground state does not exist in
the Fock space. In the generic case either the massless field must 
develop a mass or otherwise become modified by its self-interactions 
or Lorentz invariance is necessarily broken, and secular terms develop 
in the evolution from generic initial conditions. From the effective
field theory point of view, a mass and other interaction terms for a 
scalar field in two dimensions are generically allowed, and would be 
expected to control the behavior of the theory at large distances and 
late times. These additional interactions are {\it relevant} operators 
in the infrared, and cannot be treated as small perturbations to the 
massless theory. Conversely, if the masslessness of the scalar boson 
is protected by a global symmetry, then that symmetry is restored by 
quantum fluctuations and there are again no Lorentz invariant massless
Goldstone scalar states in the physical spectrum \cite{MWC,Ber}.

Infrared divergences in even classical scattering amplitudes show that 
there is no analog of a scattering matrix for gravitational waves in 
global de Sitter spacetime \cite{FIT}, much as in the two dimensional 
massless scalar theory in flat spacetime. The similar logarithmic 
behavior of the graviton propagator indicates that infrared quantum 
fluctuations of the gravitational field are important in de Sitter
space, and self-interactions or additional relevant terms in the 
effective action will control the late time behavior. This is the 
same conclusion we reached from particle creation, 
fluctuation-dissipation, and thermodynamic considerations. de Sitter 
invariance is spontaneously broken by quantum effects, and the ground 
state of the gravitational field with a cosmological term is {\it not} 
global de Sitter spacetime \cite{deSMM}.
 
The authors of refs. \cite{Wood} have performed a perturbative
analysis of long wavelength gravitational fluctuations in non de Sitter 
invariant initial states up to two-loop order. They focus on the
self-interactions of gravitons generated by nonlinearities
in the classical Einstein theory itself. This work indicates the 
presence of secular terms in the quantum stress tensor of fluctuations 
about de Sitter space, tending to decrease the effective vacuum energy 
density, consistent with our earlier considerations. The authors of \cite{MAB} 
have studied the stress tensor for long wavelength cosmological 
perturbations in inflationary models as well, and also found a backreaction 
effect of the right sign to slow inflation. Some serious technical questions
concerning the gauge invariance of these results have been raised \cite{Unr}, 
addressed \cite{AW}, and raised again recently in \cite{IWald}.
  
Even if free of technical problems the perturbative backreaction of 
\cite{Wood} and \cite{MAB} is suppressed by an effective coupling constant 
of order $\lambda^2$. This is already very small (of order $10^{-16}$) for 
an initial cosmological term at the unification scale of $10^{15}$ GeV. For 
the present value of $\lambda$ in (\ref{lmeas}) any secular backreaction 
effect of order $\lambda^2$ on vacuum energy would be completely negligible.
A much larger, essentially non-perturbative effect is needed to be relevant 
to the naturalness question of dark energy in cosmology. 

Actually, as should be clear from the previous discussion, none of the 
considerations based on particle creation, thermodynamic fluctuations or 
the infrared behavior of the gravitational fluctuations in de Sitter space 
suggest that backreaction effects can be treated in a uniform perturbative 
expansion in a small parameter like $\lambda$. On the contrary, perturbation 
theory about a state which is itself infrared singular would be expected to 
break down and require a non-perturbative resummation to capture the
dominant effects. In statistical systems perturbation theory 
certainly breaks down when there are additional infrared relevant terms 
in the low energy effective field theory, and their effects do dominate
the purely perturbative contributions. For this reason we initiated the study 
of new infrared relevant operator(s) in gravity in the conformal or trace 
sector of gravity, generated by the quantum trace anomaly. A essentially 
non-perturbative treatment of this sector indicates the existence of a 
infrared renormalization fixed point in which the cosmological term is 
driven to zero. We review and update the present status of this proposal next.

\section{Quantum Theory of the Conformal Factor}

The sensitivity to horizon scale quantum fluctuations in de Sitter spacetime
reviewed in the last section strongly suggests that there is an infrared 
relevant operator in the low energy effective theory of gravity, not 
contained in the classical Einstein-Hilbert action. This is also the 
implication of the naturalness considerations in effective field theory 
with which we began our discussion. Since there is no preferred value
of $\Lambda$ in the purely classical theory, a dynamical mechanism for the
relaxation of $\Lambda \rightarrow 0$ must be sought in the larger quantum
framework.

There are several hints for the source of new infrared relevant terms in
the quantum framework which are not contained in the classical Einstein 
theory. First it is the fluctuations in the scalar or conformal sector 
of the metric field which are the most infrared divergent in de 
Sitter spacetime \cite{AntM}. These are associated with the trace of 
the polarization tensor, and are parameterized by the conformal part 
of the metric tensor,
\begin{equation}
g_{ab}(x) = e^{2 \sigma(x)} \bar g_{ab}(x), 
\label{confdef}
\end{equation}
where $e^{\sigma}$ is called the conformal factor and $\bar g_{ab}(x)$ is 
a fixed fiducial metric. The RW scale factor $a(\tau)$ in (\ref{RWflat})
is an example of a conformal factor, fixed classically by the Friedman
eq. (\ref{rw}). The second clue that this is the important sector to look 
for non-perturbative infrared effects is that $\sigma$ couples to the 
trace of the energy-momentum tensor $T_a^{\ a}$, an operator known to 
have an anomaly for massless quantum fields in curved spacetime \cite{anom}. 
An anomaly implies that the effects of quantum fluctuations can
remain relevant at the longest length and time scales, and therefore 
modify the purely classical theory. This is certainly
the lesson of two dimensional quantum gravity, which we review next.
Most importantly of all, the scalar $\sigma$ field is constrained in 
the Einstein theory in four dimensions, but acquires {\it dynamics} 
through the trace anomaly. It is the effective action and dynamics 
of this field which we have proposed as the essential new ingredient 
to gravity at cosmological distance scales which can provide a natural 
mechanism for screening the cosmological term \cite{AntM}.

\subsection{Quantum Gravity in Two Dimensions}

Classical fields satisfying wave equations with zero mass, which are
invariant under conformal transformations of the spacetime metric, 
$g_{ab} \rightarrow e^{2\sigma} g_{ab}$ have stress tensors with zero 
classical trace, $T_a^{\ a} = 0$. In quantum theory the stress tensor
$T_a^{\ b}$ becomes an operator with fluctuations about its mean value.
The mean value itself $\langle T_a^{\ b}\rangle$ is formally UV divergent,
due to its zero point fluctuations, as in (\ref{zeropt}), and requires a 
careful renormalization procedure. The result of this renormalization
consistent with covariant conservation in curved spacetime is that
classical conformal invariance cannot be maintained at the quantum level. 
The trace of the stress tensor is generally non-zero when $\hbar \ne 0$, 
and any UV regulator which preserves the covariant conservation of $T_a^{\ b}$
(a necessary requirement of any theory respecting general coordinate invariance
and consistent with the Equivalence Principle) yields an expectation value of 
the quantum stress tensor with a non-zero trace \cite{BiD,anom}. 

In two dimensions the trace anomaly takes the simple form,
\begin{equation}
\langle T_a^{\ a} \rangle = \frac{N}{24\pi}\, R\,,\qquad (d=2)
\label{trtwo}
\end{equation}
where $N = N_S + N_F$ is the total number of massless fields, either
scalar ($N_S$) or (complex) fermionic ($N_F$). The fact that the anomalous
trace is independent of the quantum state of the matter field(s),
and dependent only on the geometry through the local Ricci scalar $R$ 
suggests that it should be regarded as a geometric effect. However, no 
local coordinate invariant action exists whose metric variation leads 
to (\ref{trtwo}). This is important because it shows immediately that 
understanding the anomalous contributions to the stress tensor will bring 
in some non-local physics or boundary conditions on the quantum state
at large distance scales. 

A non-local action corresponding to (\ref{trtwo}) can be found by 
introducing the conformal parameterization of the metric (\ref{confdef})
and noticing that the scalar curvature densities of the two metrics 
$g_{ab}$ and $\bar g_{ab}$ are related by
\begin{equation}
R \,\sqrt{-g} = \bar R \,\sqrt{-\bar g} - 2 \,\sqrt{-\bar g} \sqb \sigma\,,
\qquad (d=2)
\label{RRbar}
\end{equation}
a linear relation in $\sigma$ in two (and only two) dimensions. Multiplying 
(\ref{trtwo}) by $\sqrt{-g}$, using (\ref{RRbar}) and noting that 
$\sqrt{-g}\langle T_a^{\ a} \rangle$ defines the conformal variation,
$\delta \Gamma^{(2)}/\delta \sigma$ of an effective action $\Gamma^{(2)}$, 
we conclude that the $\sigma$ dependence of $\Gamma^{(2)}$ can be at most 
quadratic in $\sigma$. Hence the Wess-Zumino effective action \cite{WZ} in two 
dimensions, $\Gamma_{WZ}^{(2)}$ is
\begin{equation}
\Gamma_{WZ}^{(2)} [\bar g ; \sigma ] = \frac{N}{24\pi}  
\int\,d^2x\,\sqrt{-\bar g}
\left( - \sigma \sqb \sigma + \bar R\,\sigma\right)\,.
\label{WZact}
\end{equation}
Mathematically the fact that this action functional of the base metric $\bar g_{ab}$ 
and the Weyl shift parameter $\sigma$ cannot be reduced to a single local
functional of the full metric (\ref{confdef}) means that the local Weyl
group of conformal transformations has a non-trivial cohomology,
and $\Gamma_{WZ}^{(2)}$ is a one-form representative of this cohomology 
\cite{AMMc,MazMot}. This is just a formal mathematical statement of the 
fact that a effective action that incorporates the trace anomaly 
in a covariant EFT consistent with the Equivalence Principle 
must exist but that this $S_{anom}[g]$ is necessarily {\it non-local}.

It is straightforward in fact to find a non-local scalar functional 
$S_{anom}[g]$ such that
\begin{equation}
\Gamma_{WZ}^{(2)} [\bar g ; \sigma ] = S_{anom}^{(2)}[g= e^{2\sigma}\bar g]
- S_{anom}^{(2)}[\bar g]\,.
\label{cohom}
\end{equation}
By solving (\ref{RRbar}) formally for $\sigma$, and using the 
fact that $\sqrt{-g} \sq = \sqrt{-\bar g} \sqb$ is conformally invariant 
in two dimensions, we find that $\Gamma_{WZ}^{(2)}$ can be written as 
a Weyl shift (\ref{cohom}) with
\begin{equation}
S_{anom}^{(2)}[g] = \frac{Q^2}{16\pi} \int\,d^2x\,\sqrt{-g}
\int\,d^2x'\,\sqrt{-g'}\, R(x)\,{\sq}^{-1}(x,x')\,R(x')\,,
\label{acttwo}
\end{equation}
and ${\sq}^{-1}(x,x')$ denoting the Green's function inverse of the scalar 
differential operator $\sq$. The parameter $Q^2$ is $-N/6$ if only matter fields 
in a fixed spacetime metric are considered. It becomes $(25 - N)/6$ if account is 
taken of the contributions of the metric fluctuations themselves in addition to 
those of the $N$ matter fields, thus effectively replacing $N$ by $N-25$ \cite{dress}. 
In the general case, the coefficient $Q^2$ is arbitrary, related to the matter 
central charge, and can be treated as simply an additional free parameter of 
the low energy effective action, to be determined.

The anomalous effective action (\ref{acttwo}) is a scalar under coordinate 
transformations and therefore fully covariant and geometric in character,
as required by the Equivalence Principle. However since it involves the 
Green's function $\sq^{-1}(x,x')$, which requires boundary conditions for 
its unique specification, it is quite non-local, and dependent upon more 
than just the local curvature invariants of spacetime. In this important 
respect it is quite different from the classical terms in the action,
and describes rather different physics. In order to expose that physics 
it is most convenient to recast the non-local and non-single valued 
functional of the metric, $S_{anom}^{(2)}$ into a local form by 
introducing auxiliary fields. In the case of (\ref{acttwo}) a single 
scalar auxiliary field, $\varphi$ satisfying
\begin{equation}
- \sq \varphi = R
\label{auxeomtwo} 
\end{equation}
is sufficient. Then varying
\begin{equation}
S_{anom}^{(2)}[g;\varphi]  \equiv \frac{Q^2}{16\pi} \int\,d^2x\,\sqrt{-g}\,
\left(g^{ab}\,\nabla_a \varphi\,\nabla_b \varphi - 2 R\,\varphi\right)
\label{actauxtwo}
\end{equation}
with respect to $\varphi$ gives the eq. of motion (\ref{auxeomtwo})
for the auxiliary field, which when solved formally by $\varphi =-{\sq}^{-1}R$
and substituted back into $S_{anom}^{(2)}[g;\varphi]$ returns the non-local
form of the anomalous action (\ref{acttwo}), up to a surface term. 
The non-local information in addition to the local geometry which was previously 
contained in the specification of the Green's function ${\sq}^{-1}(x,x')$ 
now resides in the local auxiliary field $\varphi (x)$, and the freedom 
to add to it homogeneous solutions of (\ref{auxeomtwo}).

The variation of (\ref{actauxtwo}) with respect to the metric yields 
a stress-energy tensor,
\begin{eqnarray}
&&T_{ab}^{(2)}[g; \varphi] \equiv -\frac{2}{\sqrt{-g}} 
{\delta S_{anom}^{(2)} [g; \varphi] \over \delta g^{ab}} \nonumber\\
&& \quad = {Q^2\over 4\pi}\left[-\nabla_a\nabla_b \varphi
+ g_{ab}\, \sq\varphi - {1\over 2}(\nabla_a\varphi)(\nabla_b\varphi)
+ {1\over 4} g_{ab}\, (\nabla_c\varphi)(\nabla^c\varphi)\right]\,,
\label{anomTtwo}
\end{eqnarray}
which is covariantly conserved, by use of (\ref{auxeomtwo}) and the 
vanishing of the Einstein tensor, $G_{ab} = R_{ab} - Rg_{ab}/2 = 0$ 
in two (and only two) dimensions. The {\it classical} trace of the 
stress tensor,
\begin{equation}
g^{ab}T_{ab}^{(2)}[g; \varphi] = {Q^2\over 4\pi} \sq \varphi
=- {Q^2\over 4\pi}\,R
\label{trTtwo}
\end{equation}
reproduces the {\it quantum} trace anomaly in a general classical background
(with $Q^2$ proportional to $\hbar$). Hence (\ref{actauxtwo}) is exactly the 
local auxiliary field form of the effective action which should be added to the 
action for two dimensional gravity to take the trace anomaly of massless quantum 
fields into account.

Since the integral of $R$ is a topological invariant in two dimensions, 
the classical Einstein-Hilbert action contains no propagating degrees of freedom 
whatsoever in $d=2$, and it is $S_{anom}$ which contains the {\it only} kinetic 
terms of the low energy EFT. In the local auxiliary field form (\ref{actauxtwo}), 
it is clear that $S_{anom}$ describes an additional scalar degree of freedom 
$\varphi$, not contained in the classical action $S_{cl}^{(2)}$. This is 
reflected also in the shift of the central charge from $N-26$, which would 
be expected from the contribution of conformal matter plus ghosts by one unit 
to $N-25$. Quantum gravity in two dimensions acquires new dynamics in its
conformal sector, not present in the classical theory. Once the anomalous
term is treated in the effective action on a par with the classical terms,
its effects become non-perturbative and do not rely on fluctuations
from a given classical background to remain small.

Extensive study of the stress tensor (\ref{trTtwo}) and its correlators, arising 
from this effective action established that the two dimensional trace anomaly 
gives rise to a modification or gravitational ``dressing" of critical exponents 
in conformal field theories at second order critical points \cite{dress}. 
Since critical exponents in a second order phase transition depend only 
upon fluctuations at the largest allowed infrared scale, this dressing is 
clearly an infrared effect, independent of any ultraviolet cutoff. These 
dressed exponents are evidence of the infrared fluctuations of the 
additional scalar degree of freedom $\varphi$ which are quite absent 
in the classical action. The scaling dimensions of correlation functions 
so obtained are clearly non-perturbative in the sense that they are not 
obtained by considering perturbatively small fluctuations around flat space, 
or controlled by a uniform expansion in $\lambda \ll 1$. The appearance 
of the gravitational dressing exponents and the anomalous effective action 
(\ref{acttwo}) itself have been confirmed in the large volume scaling 
limit of two dimensional simplicial lattice simulations in the dynamical 
triangulation approach \cite{DT,CatMot}. Hence there can be little doubt
that the anomalous effective action (\ref{actauxtwo}) correctly accounts
for the infrared fluctuations of two dimensional geometries.

The importance of this two dimensional example is the lessons it allows
us to draw about the role of the quantum trace anomaly in the low energy
EFT of gravity, and in particular the new dynamics it contains in the
conformal factor of the metric. The effective action generated by the 
anomaly in two dimensions contains a {\it new} scalar degree of freedom, 
relevant for infrared physics, beyond the purely local classical action. 
It is noteworthy that the new scalar degree of freedom in (\ref{auxeomtwo}) 
is massless, and hence fluctuates at all scales, including the very
largest allowed. In two dimensions its propagator ${\sq}^{-1}(x,x')$ 
is logarithmic, as in (\ref{logtwo}), and hence is completely unsuppressed 
at large distances. Physically this means that the quantum correlations
at large distances require additional long wavelength information such
as macroscopic boundary conditions on the quantum state.

The action (\ref{actauxtwo}) due to the anomaly is exactly the missing 
relevant term in the low energy EFT of two dimensional gravity responsible 
for non-perturbative fluctuations at the largest distance scales. This 
modification of the classical theory is {\it required} by general covariance 
and quantum theory, and  essentially {\it unique} within the EFT framework. 

\section{Quantum Conformal Factor in Four Dimensions}

The line of reasoning in $d=2$ dimensions just sketched to find the 
conformal anomaly and construct the effective action may be followed 
also in four dimensions. In $d=4$ the trace anomaly takes the 
somewhat more complicated form,
\begin{equation}
\langle T_a^{\ a} \rangle
= b F + b' \left(E - \frac{2}{3}\sq R\right) + b'' \sq R + \sum_i \beta_iH_i\,,
\label{tranom}
\end{equation}
in a general four dimensional curved spacetime. This is the four
dimensional analog of (\ref{trtwo}) in two dimensions.
In eq. (\ref{tranom}) we employ the notation,
\numparts
\begin{eqnarray}
&&E \equiv ^*\hskip-.2cmR_{abcd}\,^*\hskip-.1cm R^{abcd} = 
R_{abcd}R^{abcd}-4R_{ab}R^{ab} + R^2 
\,,\qquad {\rm and} \label{EFdef}\\
&&F \equiv C_{abcd}C^{abcd} = R_{abcd}R^{abcd}
-2 R_{ab}R^{ab}  + {R^2\over 3}\,.
\end{eqnarray}
\endnumparts

\noindent  with $R_{abcd}$ the Riemann curvature tensor, 
$^*\hskip-.1cmR_{abcd}= \varepsilon_{abef}R^{ef}_{\ \ cd}/2$ its dual, 
and $C_{abcd}$ the Weyl conformal tensor. Note that $E$ is the four 
dimensional Gauss-Bonnet combination whose integral gives the Euler 
number of the manifold, analogous to the Ricci scalar $R$ in $d=2$.
The coefficients $b$, $b'$ and $b''$ are dimensionless parameters multiplied
by $\hbar$. Additional terms denoted by the sum $\sum_i \beta_i H_i$ in 
(\ref{tranom}) may also appear in the general form of the trace anomaly, 
if the massless conformal field in question couples to additional long range 
gauge fields. Thus in the case of massless fermions coupled to a background 
gauge field, the invariant $H =$tr($F_{ab}F^{ab}$) appears in (\ref{tranom}) 
with a coefficient $\beta$ determined by the anomalous dimension of the 
relevant gauge coupling. 

As in $d=2$ the form of (\ref{tranom}) and the coefficients $b$ and $b'$ are 
independent of the state in which the expectation value of the stress tensor 
is computed, nor do they depend on any ultraviolet short distance cutoff. 
Instead their values are determined only by the number of massless fields 
\cite{BiD,anom}, 
\numparts
\begin{eqnarray}
b &=& \frac{1}{120 (4 \pi)^2}\, (N_S + 6 N_F + 12 N_V)\,,\\
b'&=& -\frac{1}{360 (4 \pi)^2}\, (N_S + {11\over 2} N_F + 62 N_V)\,,
\label{bprime}
\end{eqnarray}
\label{bbprime}
\endnumparts
\vskip -.3cm
\noindent with $(N_S, N_F, N_V)$ the number of fields of spin 
$(0, \frac{1}{2}, 1)$ respectively and we have taken $\hbar = 1$.
Notice also that $b >0$ while $b' < 0$ for all fields of lower spin 
for which they have been computed. Hence the trace anomaly can lead 
to stress tensors of either sign, and in particular, of the sign
needed to compensate for a positive bare cosmological term. The 
anomaly terms can also be utilized to generate an effective positive 
cosmological term if none is present initially. Such anomaly driven 
inflation models \cite{Starob} require curvatures comparable to the 
Planck scale, unless the numbers of fields in (\ref{bbprime}) is 
extremely large. It is clear that conformally flat cosmological models 
of this kind, in which the effects of the anomaly can be reduced to a 
purely local higher derivative stress tensor, are of no relevance to the 
very small cosmological term (\ref{cosmeas}) we observe in the acceleration 
of the Hubble expansion today. Instead it is the essentially {\it non-local} 
boundary effects of the anomaly on the horizon scale, much larger than 
$L_{Pl}$ which should come into play.

Three local fourth order curvature invariants $E, F$ and $\sq R$ appear in 
the trace of the stress tensor (\ref{tranom}), but only the first two 
(the $b$ and $b'$) terms of (\ref{tranom}) cannot be derived from a local 
effective action of the metric alone. If these terms could be derived from 
a local gravitational action we would simply make the necessary finite 
redefinition of the corresponding local counterterms to remove them from 
the trace, in which case the trace would no longer be non-zero or anomalous. 
This redefinition of a local counterterm (namely, the $R^2$ term in the 
effective action) is possible only with respect to the third $b''$ coefficient 
in (\ref{tranom}), which is therefore regularization dependent and not part of 
the true anomaly. Only the non-local effective action corresponding to the 
$b$ and $b'$ terms in (\ref{tranom}) are independent of the UV regulator and 
lead to effects that can extend over arbitrarily large, macroscopic distances. 
The distinction of the two kinds of terms in the effective action is emphasized 
in the cohomological approach to the trace anomaly \cite{MazMot}.

The number of massless fields of each spin ($N_S, N_F, N_V$) is a property of 
the low energy effective description of matter, having no direct connection 
with physics at the ultrashort Planck scale. Indeed massless fields fluctuate 
at all distance scales and do not decouple in the far infrared, relevant for
cosmology. As in the case of the chiral anomaly with massless quarks, the 
$b$ and $b'$ terms in the trace anomaly were calculated originally by 
techniques usually associated with UV regularization \cite{BiD}. However 
just as in the case of the chiral anomaly in QCD, or two dimensional gravity, 
the trace anomaly can have significant new infrared effects, not captured
by a purely local metric description.

To find the WZ effective action corresponding to the $b$ and $b'$ terms in
(\ref{tranom}), introduce as in two dimensions the conformal parameterization 
(\ref{confdef}), and compute
\numparts
\begin{eqnarray}
&&\sqrt{-g}\,F = \sqrt{-\bar g}\,\bar F\,\label{Fsig}\\
&&\sqrt{-g}\,\left(E - {2\over 3}\sq R\right) = \sqrt{-\bar g}\,
\left(\overline E - {2\over 3}\sqb\overline R\right) + 4\,\sqrt{-\bar g}\,
\bar\Delta_4\,\sigma\,,
\label{Esig}
\end{eqnarray}
\label{FEsig}
\endnumparts
\vskip-.3cm
\noindent whose $\sigma$ dependence is no more than linear. The 
fourth order differential operator appearing in this expression is 
\cite{AntM,MazMot,Rie}
\begin{equation}
\Delta_4 \equiv \sq^2 + 2 R^{ab}\nabla_a\nabla_b - {2\over 3} R \sq + 
{1\over 3} (\nabla^a R)\nabla_a \,,
\label{Deldef}
\end{equation}
which is the unique fourth order scalar operator that is conformally covariant, 
{\it viz.}
\begin{equation}
\sqrt{-g}\, \Delta_4 = \sqrt{-\bar g}\, \bar \Delta_4 \,,
\label{invfour}
\end{equation}
for arbitrary smooth $\sigma(x)$ in four (and only four) dimensions. 
Thus multiplying (\ref{tranom}) by $\sqrt{-g}$ and recognizing that the
result is the $\sigma$ variation of an effective action $\Gamma_{WZ}$, we
find immediately that this quadratic effective action is
\begin{equation}
\hspace{-2.2cm}
\Gamma_{WZ}[\bar g;\sigma] = b  \int\,d^4x\,\sqrt{-\bar g}\, \bar F\,\sigma
+ b' \int\,d^4x\,\sqrt{-\bar g}\,\left\{\left(\bar E - {2\over 3}
\sqb \bar R\right)\sigma + 2\,\sigma\bar\Delta_4\sigma\right\}\,,
\label{WZfour}
\end{equation}
up to terms independent of $\sigma$. This Wess-Zumino action is a 
one-form representative of the non-trivial cohomology of the local 
Weyl group in four dimensions which now contains two distinct cocycles, 
corresponding to the two independent terms multiplying $b$ and $b'$. 
By solving (\ref{Esig}) formally for $\sigma$ and substituting
the result in (\ref{WZfour}) we obtain
\begin{equation}
\Gamma_{WZ}[\bar g;\sigma] = S_{anom}[g=e^{2\sigma}\bar g] - S_{anom}[\bar g],
\label{Weylshift}
\end{equation}
with the {\it non-local} anomalous action,
\begin{eqnarray}
&&\hspace{-2.2cm}
S_{anom}[g] =  {1\over 2}\int d^4x\sqrt{g}\int d^4x'\sqrt{g'}\,
\left(\frac{E}{2} - \frac{\sq R}{3}\right)_x \Delta_4^{-1} (x,x')\left[ bF + b'
\left(\frac {E}{2} - \frac{\sq R}{3}\right)\right]_{x'}\nonumber\\
\label{nonl}
\end{eqnarray}
and $\Delta_4^{-1}(x,x')$ denoting the Green's function inverse of the fourth 
order differential operator defined by (\ref{Deldef}). From the foregoing
construction it is clear that if there are additional Weyl invariant
terms in the anomaly (\ref{tranom}) they should be included in the
$S_{anom}$ by making the replacement $bF \rightarrow bF + \sum_i\beta_i H_i$
in the last square bracket of (\ref{nonl}).

Notice from the derivation of $S_{anom}$ that although the $\sigma$
independent piece of the gravitational action cannot be determined
from the trace anomaly alone, the $\sigma$ dependence is {\it uniquely}
determined by the general form of the trace anomaly for massless
fields. Thus, whatever else may be involved in the full quantum theory
of gravity in four dimensions at short distance scales, the anomalous
effective action (\ref{nonl}) should be included in the gravitational
action at large distance scales, {\it i.e.} in the far infrared.
Again the physics is that the quantum fluctuations of massless fields 
do not decouple and contribute to gravitational effects at arbitrarily 
large distances. Graviton ({\it i.e.} spin-two) fluctuations of the metric 
should give rise to an effective action of precisely the same form as 
$S_{anom}$ with new coefficients $b$ and $b'$, which can be checked at 
one-loop order \cite{AMMc}. The effective action (\ref{nonl}) or 
(\ref{flata}) is the starting point for studying the new physics of 
quantum fluctuations of the conformal factor and infrared renormalization 
in gravity which allows the bare cosmological term of the classical 
Einstein theory to be screened. We postpone until Sec. 7 the discussion
of the auxiliary field description of (\ref{nonl}) and the full
low energy effective field theory of four dimensional gravity.

\section{Finite Volume Scaling and Infrared Screening of $\lambda$}

Let us consider the dynamical effects of the anomalous terms in the 
simplest case that the fiducial metric is conformally flat, {\it i.e.}
$g_{ab} = e^{2 \sigma}\eta_{ab}$. Then the Wess Zumino effective action 
simplifies to
\begin{eqnarray}
\Gamma_{WZ} [\eta; \sigma] = -\frac{Q^2}{16\pi^2} \int d^4 x \ (\sq \sigma)^2 \,,
\label{flata}
\end{eqnarray}
where 
\begin{equation}
Q^2 \equiv -32 \pi^2 b'\,.
\label{Qdef}
\end{equation}
This action quadratic in $\sigma$
is the action of a free scalar field, albeit with a kinetic term that 
is fourth order in derivatives. The propagator for this kinetic term is 
$(p^2)^{-2}$ in momentum space, which is a logarithm in position space, 
\begin{equation}
G_{\sigma} (x,x') = -\frac{1}{2Q^2} \, \ln \left[\mu^2 (x-x')^2\right]
\end{equation}
the same as (\ref{logtwo}) in two dimensions. Of course this is no accident 
but rather a direct consequence of the association with the anomaly
of a conformally invariant differential operator, $\sq$ in two dimensions
and $\Delta_4$ in four dimensions, a pattern which continues
in all higher even dimensions. Because of this logarithmic propagator
we must expect the similar sort of infrared fluctuations, conformal fixed 
point and dressing exponents as those obtained in two dimensional gravity.

The classical Einstein-Hilbert action for a conformally flat
metric $g_{ab} = e^{2 \sigma}\eta_{ab}$ is
\begin{equation}
\frac{1}{8\pi G}\ \int d^4 x \ \left[ 3 e^{2 \sigma} 
(\partial_a \sigma)^2 - \Lambda e^{4 \sigma} \right]\,,
\label{cEin}
\end{equation}
which has derivative and exponential self-interactions in $\sigma$.
It is remarkable that these complicated interactions can be treated 
systematically using the the fourth order kinetic term of (\ref{flata}). 
In fact, these interaction terms are renormalizable and their anomalous 
scaling dimensions due to the fluctuations of $\sigma$ can be computed 
in closed form \cite{AntM,AMMd}. Direct calculation of the conformal 
weight of the Einstein curvature term shows that it acquires an 
anomalous dimension $\beta_2$ given by the quadratic relation,
\begin{equation}
\beta_2 = 2 + \frac{\beta_2^2}{2Q^2}\,.
\label{betE}
\end{equation}
In the limit $Q^2 \rightarrow \infty$
the fluctuations of $\sigma$ are suppressed and we recover the
classical scale dimension of the coupling $G^{-1}$ with mass
dimension $2$. Likewise the cosmological term in (\ref{cEin}) corresponding 
to the four volume acquires an anomalous dimension given by 
\begin{equation}
\beta_0 = 4 + \frac{\beta_0^2}{2Q^2}\,.
\label{betL}
\end{equation}
Again as $Q^2 \rightarrow \infty$ the effect of the fluctuations of
the conformal factor are suppressed and we recover the classical scale
dimension of $\Lambda/G$, namely four. The solution of the quadratic
relations (\ref{betE}) and (\ref{betL}) determine the scaling dimensions 
of these couplings at the conformal fixed point at other values of $Q^2$. 
This can be extended to local operators of any non-negative integer
mass dimension $p$, with associated couplings of mass dimension $4-p$, by 
\begin{equation}
\beta_p = 4-p + \frac{\beta_p^2}{2Q^2}\,.
\label{betn}
\end{equation}
In order to obtain the classical scale dimension $4-p$ in the limit
$Q^2 \rightarrow \infty$ the sign of the square root is determined
so that
\begin{equation}
\beta_p = Q^2 \left[ 1 - \sqrt{1 - \frac{(8-2p)}{Q^2}}\right]\,,
\end{equation}
valid for $Q^2 \ge 8 - 2p$ for all $p\ge 0$, and thus $Q^2 \ge 8$. 
These scaling dimensions were computed both by covariant and canonical
operator methods. In the canonical method we also showed that the anomalous 
action for the conformal factor does not have unphysical ghost or tachyon
modes in its spectrum of physical states \cite{AMM}.

In the language of statistical mechanics and critical phenomena the
quadratic action (\ref{flata}) describes a Gaussian conformal fixed 
point, where there are no scales and conformal invariance is
exact. The positive corrections of order $1/Q^2$ (for $Q^2 > 0$)
in (\ref{betE}) and (\ref{betL}) show that this fixed point is
stable in the infrared, that is, both couplings $G^{-1}$ and
$\Lambda/G$ flow to zero at very large distances. Because both 
of these couplings are separately dimensionful, at a conformal fixed
point one should properly speak only of the dimensionless
combination $\hbar G \Lambda/ c^3 = \lambda$. By normalizing
to a fixed four volume $V= \int d^4 x$ one can show that the finite
volume renormalization of $\lambda$ is controlled by the anomalous 
dimension,
\begin{equation}
2 \delta - 1 \equiv 2\, \frac{\beta_2}{\beta_0} - 1 = 
\frac{\sqrt{1 - {8\over Q^2}} - \sqrt{1 - {4\over Q^2}}}
{1 + \sqrt{1 - {4\over Q^2}}} \le 0\,.
\label{scal}
\end{equation}
This is the anomalous dimension that enters the infrared
renormalization group volume scaling relation \cite{AMMc},
\begin{equation}
V \frac{d}{d V} \lambda = 4\, (2 \delta - 1)\, \lambda\,.
\label{renL}
\end{equation}
The anomalous scaling dimension (\ref{scal}) is negative for all 
$Q^2 \ge 8$, starting at $1 - \sqrt 2 = -0.414$ at $Q^2 =8$
and approaching zero as $- 1/Q^2$ as $Q^2 \rightarrow \infty$. 
This implies that the dimensionless cosmological term $\lambda$ has 
an infrared fixed point at zero as $V\rightarrow \infty$. Thus the 
cosmological term is dynamically driven to zero as $V\rightarrow \infty$
by infrared fluctuations of the conformal part of the metric described 
by (\ref{flata}).

We emphasize that no fine tuning is involved here and no
free parameters enter except $Q^2$, which is determined by the trace
anomaly coefficient $b'$ by (\ref{Qdef}). Once $Q^2$ is assumed
to be positive, then $2 \delta - 1$ is negative, and $\lambda$ is
driven to zero at large distances by the conformal fluctuations 
of the metric, with no additional assumptions. 

The result (\ref{renL}) does rely on the use of (\ref{flata}) or its 
curved space generalization (\ref{nonl}) as the free kinetic term in the 
effective action for gravity, treating the usual Einstein-Hilbert terms as 
interactions or ``marginal deformations" of the conformal fixed point. This 
conformal fixed point represents a new phase of gravity, non-perturbative 
in any expansion about flat space. In this phase conformal invariance
is restored and the mechanism of screening $\lambda$ due to quantum effects 
proposed in \cite{AntM} is realized. 

Identifying the fluctuations responsible for driving $\lambda$ to zero 
within a framework based on quantum field theory and the Equivalence Principle, 
free of {\it ad hoc} assumptions or fine tuning is an important first step
towards a full solution of the cosmological constant problem. However, the 
application of this screening mechanism to cosmology, in which we presume a 
classical or semiclassical line element of the form (\ref{RWflat}), is unclear.
Near the conformal fixed point the inverse Newtonian constant $G^{-1}$ is
also driven to zero when compared to some fixed mass scale $m$ \cite{AMMd}.
This is clearly different from the situation we observe in our local
neighborhood.  Under what conditions and where exactly (\ref{nonl}) can
dominate the classical Einstein terms, and moreover how the screening 
mechanism could be used to relax the vacuum energy to zero in a realistic 
cosmological model are questions not answered by our considerations to this 
point. Absent such a complete theory of cosmological vacuum energy we 
proposed that the conformally invariant phase might be relevant on horizon 
scales in cosmology. In that case the signatures of conformal invariance 
should be imprinted on and observable in the spectrum and statistics of the CMB.

\section{Conformal Invariance and the CMB} 

Our earlier studies of fluctuations in de Sitter space suggest
that the fluctuations responsible for the screening of $\lambda$ take 
place at the horizon scale. In that case then the microwave photons 
in the CMB reaching us from their surface of last scattering should 
retain some imprint of the effects of these fluctuations. It then 
becomes natural to extend the classical notion of scale invariant 
cosmological perturbations, pioneered by Harrison and Zel'dovich \cite{HZ} 
to full conformal invariance. In that case the classical spectral index 
of the perturbations should receive corrections due to the anomalous 
scaling dimensions at the conformal phase \cite{sky}. In addition to 
the spectrum, the statistics of the CMB should reflect the non-Gaussian 
correlations characteristic of conformal invariance. This generic
dynamical prediction of non-Gaussian correlations in the CMB due to conformal
invariance was made for the first time to our knowledge in ref. \cite{sky}.

Scale invariance was introduced into physics in early attempts to describe the 
apparently universal behavior observed in turbulence and second order phase 
transitions, which are independent of the particular short distance dynamical 
details of the system. The gradual refinement and development of this simple 
idea of universality led to the modern theory of critical phenomena, one of 
whose hallmarks is well-defined logarithmic deviations from naive scaling 
relations \cite{RG}. A second general feature of the theory is the specification 
of higher point correlation functions of fluctuations according to 
the requirements of conformal invariance at the critical point \cite{pol}. 

In the language of critical phenomena, the observation of Harrison and 
Zel'dovich that the primordial density fluctuations should be characterized by 
a spectral index $n=1$ is equivalent to the statement that the observable giving 
rise to these fluctuations has engineering or naive scaling dimension $p = 2$. 
This is because the density fluctuations $\delta \rho$ are related to the metric 
fluctuations by Einstein's equations, $\delta R \sim G\delta\rho$, which is second 
order in derivatives of the metric. Hence, the two-point spatial correlations 
$\langle \delta\rho(x) \delta\rho(y)\rangle \sim \langle \delta R(x) \delta R(y)\rangle$
should behave like $|x-y|^{-4}$, or $|k|^1$ in Fourier space, according to simple 
dimensional analysis.

One of the principal lessons of the modern theory of critical phenomena 
is that the transformation properties of observables under conformal transformations 
at the fixed point is {\it not} given by naive dimesnional analysis. 
Rather one should expect to find well-defined logarithmic deviations 
from naive scaling, corresponding to a (generally non-integer) dimension 
$\Delta \ne p$. The deviation from naive scaling $\Delta - p$ 
is the ``anomalous'' dimension of the observable due to critical fluctuations, 
which may be quantum or statistical in origin. Once $\Delta$ is fixed for a 
given observable, the requirement of conformal invariance determines the form 
of its two- and three-point correlation functions up to an arbitrary amplitude, 
without reliance on any particular dynamical model. 

Consider first the two-point function of any observable ${\cal O}_{\Delta}$
with dimension $\Delta$. Conformal invariance requires \cite{RG,pol}
\begin{equation}
\langle{\cal O}_{\Delta} (x_1) {\cal O}_{\Delta} (x_2)\rangle
\sim \vert x_1-x_2 \vert^{-2\Delta}
\end{equation}
at equal times in three dimensional flat spatial coordinates. In
Fourier space this gives
\begin{equation}
G_2(k) \equiv\langle\tilde{\cal O}_{\Delta} (k) \tilde 
{\cal O}_{\Delta} (-k)\rangle \sim \vert k \vert^{2\Delta - 3} \,.
\label{G2}
\end{equation}
Thus, we define the spectral index of this observable by
\begin{equation}
n \equiv 2 \Delta - 3\ .
\label{index}
\end{equation}
In the case that the observable is the primordial density fluctuation 
$\delta\rho$, and in the classical limit where its anomalous dimension 
vanishes, $\Delta \rightarrow p =2$, we recover the 
Harrison-Zel'dovich spectral index of $n=1$.

In order to convert the power spectrum of primordial density 
fluctuations to the spectrum of fluctuations in the CMB at large angular 
separations we follow the standard treatment \cite{peeb} relating the
temperature deviation to the Newtonian gravitational potential 
$\varphi$ at the last scattering surface, ${\delta T \over T} \sim \delta 
\varphi$, which is related to the density perturbation in turn by 
\begin{equation}
\nabla^2 \delta\varphi = 4\pi G\, \delta\rho \ .
\label{lap}
\end{equation}
Hence, in Fourier space, 
\begin{equation}
{\delta T \over T} \sim \delta \varphi \sim 
{1\over k^2}{\delta\rho\over \rho}\ ,
\end{equation}
and the two-point function of CMB temperature fluctuations is
determined by the conformal dimension $\Delta$ to be
\begin{eqnarray}
&&C_2(\theta) \equiv \left\langle{\delta T \over T}(\hat r_1)
{\delta T \over T}(\hat  r_2)\right\rangle
\sim \nonumber\\
&&\int d^3 k\left({1\over k^2}\right)^2 G_2(k) e^{i k\cdot 
r_{12}}
\sim \Gamma (2-\Delta) (r_{12}^2)^{2 - \Delta}\ ,
\label{C2}
\end{eqnarray}
where $r_{12} \equiv (\hat r_1 - \hat r_2)r$
is the vector difference between the two positions from which the 
CMB photons originate. They are at equal distance $r$ from the observer 
by the assumption that the photons were emitted at the last scattering 
surface at equal cosmic time. Since $r_{12}^2 = 2 (1- \cos \theta)r^2$, 
we find then
\begin{equation}
C_2(\theta) \sim \Gamma (2-\Delta) (1-\cos\theta)^{2 - \Delta}
\label{Ctheta} 
\end{equation}
for arbitrary scaling dimension $\Delta$.  

Expanding the function $C_2(\theta)$ in multipole moments,
\begin{equation}
C_2(\theta) = {1\over 4\pi} \sum_{\ell} (2\ell + 1)
c_{\ell}^{(2)}(\Delta) P_{\ell} (\cos \theta)\ ,
\label{c2m}
\end{equation}
\begin{equation}
c_{\ell}^{(2)}(\Delta) \sim \Gamma(2-\Delta) \sin\left[ \pi 
(2-\Delta)\right]
{\Gamma (\ell - 2 + \Delta) \over \Gamma (\ell + 4 -\Delta)}\ ,
\end{equation}
shows that the pole singularity at $\Delta =2$ appears only in the $\ell = 
0$ monopole moment. This singularity is just the reflection of the fact 
that the Laplacian in (\ref{lap}) cannot be inverted on constant functions, 
which should be excluded. Since the CMB  anisotropy is defined by removing 
the isotropic monopole moment (as well as the dipole moment), the $\ell =0$ 
term does not appear in the sum, and the higher moments of the anisotropic 
two-point correlation function are well-defined for $\Delta$ near $2$. 
Normalizing to the quadrupole moment $c_2^{(2)}(\Delta)$, we find
\begin{equation}
c_{\ell}^{(2)}(\Delta) = c_2^{(2)}(\Delta) 
{\Gamma (6 - \Delta) \over \Gamma (\Delta) } 
{\Gamma (\ell - 2 + \Delta) \over \Gamma(\ell + 4 - \Delta)}\ ,
\end{equation}
which is a standard result \cite{peeb}. Indeed, if $\Delta$ is replaced by 
$p = 2$ we obtain $\ell (\ell + 1)  c_{\ell}^{(2)}(p) = 6 c_2^{(2)} (p)$, 
which is the well-known  predicted behavior of the lower moments ($\ell \le 30 $) 
of the CMB anisotropy where the Sachs-Wolfe effect should dominate.

If the conformal fixed point behavior described in the previous section
dominates at these scales then the scaling dimension of an observable
with classical dimension $p$ is given by \cite{spike}
\begin{equation}
\Delta_p  =  4 \ \frac{\sqrt{1 - \frac{(8-2p)}{Q^2}} - \sqrt{1 - \frac{8}{Q^2}}}
{1 - \sqrt{1 - \frac{8}{Q^2}}}\ .
\label{dress}
\end{equation}
Hence consideration of the trace anomaly generated by the zero-point 
fluctuations of massless fields leads necessarily to well-defined quantum 
corrections to the naive scaling dimensions of observables in cosmology. 
In the limit $Q^2 \rightarrow \infty$, the effects of fluctuations in the 
metric are suppressed and one recovers the classical scaling dimension $p$,
\begin{equation}
\Delta_p = p + \frac{1}{2 Q^2}\, p\,(4-p) + \cdots 
\end{equation}
The quantity $Q^{-2}$ is determined in principle by the trace
anomaly coefficient $b'$ through (\ref{tranom}) and (\ref{Qdef}),
but we may regard it as simply a free parameter characterizing 
the universality class of the conformal metric fluctuations, 
which should be determined from the observations. From this slightly 
more general perspective, the conformal invariance considerations that 
lead to (\ref{dress}) are quite independent of any particular model of 
their origin. 
  
In the analysis of physical observables in the conformal sector of gravity, 
the operator with the lowest non-trivial scaling dimension corresponds, 
in the classical limit, to the scalar curvature $R$ with $p = 2$ \cite{spike}. 
Since the fluctuations which dominate at large distances correspond to 
observables with lowest scaling dimensions, the conformal factor theory 
in this limit selects precisely Harrison's original choice. With 
$p = 2$, we find a definite prediction for deviations from a strict 
Harrison-Zel'dovich spectrum according to Eqns.~(\ref{index}) and (\ref{dress}) 
in terms of the parameter $Q^2$. The resulting spectral index is always greater 
than unity for all finite $Q^2 \ge 8$, approaching one as 
\begin{equation}
n = 1 + {4\over Q^2}+\cdots
\label{indexQ}
\end{equation}
as $Q^2 \rightarrow \infty$.

The latest WMAP three year CMB results now favor a spectral index
for scalar perturbations of about $0.95$, some three standard
deviations lower than unity. From (\ref{tranom}) and (\ref{Qdef}), 
the value of $Q^2$ for free conformally invariant fields is 
\begin{equation}
Q^2 = {1 \over 180}(N_S + {\textstyle{11\over 2}} N_{F} + 62 N_V - 28) + 
Q^2_{grav}\ ,
\label{cent}
\end{equation}
where $Q^2_{grav}$ is the contribution of spin-$2$ gravitons. 
The $-28$ contribution is that of the conformal or spin-$0$ 
part of the metric itself. The main theoretical uncertainty in
determining $Q^2_{grav}$ is that the Einstein theory is neither 
conformally invariant nor free, so that a method for evaluating the  
infrared effects of spin-$2$ gravitons is required which is insensitive 
to ultraviolet physics. A purely one-loop computation gives 
$Q^2_{grav}\simeq 7.9$ for the graviton contribution \cite{AMMc}. 
Taking this estimate at face value and including all known 
fields of the Standard Model of particle physics (for which
$N_{F}=45$ and $N_V = 12$) we find
\begin{equation}
Q^2_{SM} \simeq 13.2 \qquad {\rm and} \qquad n\simeq 1.45\ ,
\label{standmod}
\end{equation}
which is now firmly excluded by the WMAP data. If we require
that $n$ be within $0.05$ of unity, then $Q^2 > 80$ is needed.

It is possible that the effective number of massless degrees of
freedom was much higher at the surface of last scattering from
which the CMB was emitted, leading to a much higher value of
$Q^2$. It is also to be noted that the same conformal fixed point 
fluctuations that led to (\ref{index}) and (\ref{dress}) also drive 
the cosmological term $\lambda$ and the inverse Newtonian constant
to zero. These non-classical effects on the geometry have 
not been taken into account in the essentially classical calculation of
(\ref{lap})-(\ref{Ctheta}). The scaling relations at the conformal 
fixed point were also derived in a four dimensional Euclidean theory. Yet 
in (\ref{lap})-(\ref{Ctheta}) we assumed spatially flat FLRW three 
dimensional sections. The actual geometry through which the CMB photons 
propagate from their last scattering surface to us may be different. 
Finally, the cosmological parameters are also extracted from the data 
through the use of models such as $\Lambda$ CDM or variants
thereof. Some of the assumptions in these models may have to be
re-examined if gravity itself is modified by the anomalous terms.
Accounting for these possible effects on the spectral index
requires a more complete cosmological model.

Turning now from the two-point function of CMB fluctuations to higher point 
correlators, we find a second characteristic prediction of conformal invariance, 
namely non-Gaussian statistics for the CMB. The first correlator sensitive to this 
departure from gaussian statistics is the three-point function of the observable 
${\cal O}_{\Delta}$, which takes the form \cite{pol}
\begin{equation}
\langle{\cal O}_{\Delta} (x_1) {\cal O}_{\Delta} (x_2) 
{\cal O}_{\Delta} (x_3)\rangle
\sim |x_1-x_2|^{-\Delta} |x_2- x_3|^{-\Delta} |x_3 - x_1|^{-\Delta}\ ,
\end{equation} 
or in Fourier space \cite{sky},
\begin{eqnarray}
&&\hskip-2.5cm G_3 (k_1, k_2) \sim \int d^3 p\ |p|^{\Delta -3}\,  |p + k_1|^{\Delta -3}\, 
|p- k_2|^{\Delta -3}\, \sim \frac{\Gamma\left(3-{\Delta\over 2}\right)}
{[\Gamma\left(\frac{9-3\Delta}{2}\right)]^3} 
\int_0^1\int_0^1\,du\,dv\times 
\nonumber\\
&&\hskip-2cm [u(1-u)v]^{\frac{1-\Delta}{2}}(1-v)^{\frac{\Delta}{2}-1}
[u(1-u)(1-v)k_1^2 + v(1-u)k_2^2 + uv(k_1+k_2)^2]^{-(3-\frac{\Delta}{2})}.
\label{three}
\end{eqnarray}
This three-point function of primordial density fluctuations gives rise to
three-point correlations in the CMB by reasoning precisely analogous as that
leading from Eqns.~(\ref{G2}) to (\ref{C2}). That is,
\begin{eqnarray}
&& C_3(\theta_{12}, \theta_{23}, \theta_{31}) 
\equiv \left\langle{\delta T \over T}(\hat r_1)
{\delta T \over T}(\hat  r_2){\delta T \over T}(\hat  
r_3)\right\rangle \nonumber \\
&& \qquad\qquad\qquad \sim \int {d^3 k_1\,d^3k_2 \over
k_1^2\, k_2^2\, (k_1 + k_2)^2}\ G_3 (k_1, k_2)\, e^{i k_1\cdot r_{13}} 
e^{ik_2\cdot r_{23}}
\label{C3}
\end{eqnarray}
where $r_{ij}\equiv ({\hat r}_i-{\hat r_j})r$ and 
$r_{ij}^2=2(1-\cos\theta_{ij}) r^2$. 

From (\ref{three}) and (\ref{C3}), it is easy to extract the global 
scaling of the three-point function, 
\begin{eqnarray}
&& G_3(s k_1, s k_2) \sim s^{3(\Delta -2)} G_3(k_1,k_2)\ ,
\nonumber\\
&& \qquad C_3 \sim r^{3(2-\Delta)}\ .
\label{gloscal}
\end{eqnarray}
In the general case of three different angles, the expression for 
the three-point correlation function (\ref{C3}) is quite complicated, 
although it can be rewritten in parametric form analogous to (\ref{three})
to facilitate numerical evaluation, if desired. An estimate of its angular 
dependence in the limit $\Delta\to 2$ can be obtained by replacing the slowly
varying $G_3(k_1,k_2)$ by a constant. Then (\ref{C3}) can be 
expanded in terms of spherical harmonics,
\begin{eqnarray}
&&\hskip-2cm C_3(\theta_{ij})\sim
\sum_{l_i,m_i}{{\cal K}^*_{l_1m_1l_2m_2l_3m_3}\over
(2l_1+1)(2l_2+1)(2l_3+1)} \times \nonumber\\
&& \left( {1\over l_1+l_2+l_3}+
{1\over l_1+l_2+l_3+3}\right) Y_{l_1m_1}({\hat r}_1) 
Y_{l_2m_2}({\hat r}_2)
Y_{l_3m_3}({\hat r}_3)\ ,
\end{eqnarray}
where ${\cal K}_{l_1m_1l_2m_2l_3m_3}\equiv\int d{\hat r} 
Y_{l_1m_1}({\hat r}) Y_{l_2m_2}({\hat r}) Y_{l_3m_3}({\hat r})$.

In the special case of equal angles $\theta_{ij}=\theta$, it follows from 
(\ref{gloscal}) that the three-point correlator is 
\begin{equation}
C_3(\theta)\sim (1-\cos\theta)^{{3\over 2}(2-\Delta)}\ .
\end{equation}
Expanding the function $C_3(\theta)$ in multiple moments as in 
Eqn.~(\ref{c2m})
with coefficients $c_{\ell}^{(3)}$, and normalizing to the quadrupole moment,
we find
\begin{equation}
c_{\ell}^{(3)}(\Delta) =c_{2}^{(3)}(\Delta)
{\Gamma (4+{3\over 2}(2-\Delta))\over\Gamma (2-{3\over 
2}(2-\Delta))}
{\Gamma (\ell-{3\over 2}(2-\Delta))\over\Gamma(\ell+2+{3\over 
2}(2-\Delta))}\ .
\label{cl3}
\end{equation}
In the limit $\Delta \rightarrow 2$, we obtain 
$\ell(\ell+1)c_{\ell}^{(3)}=6c_2^{(3)}$, which is the same result as for 
the moments $c_{\ell}^{(2)}$ of the two-point correlator but with a different 
quadrupole amplitude. 

The value of this quadrupole normalization $c_2^{(3)}(\Delta)$ cannot be 
determined by conformal symmetry considerations alone. A naive
comparison with the two-point function which has a small amplitude of 
the order of $10^{-5}$ leads to a rough estimate of $c_2^{(3)}\sim{\cal
O}(10^{-7.5})$, which would make it very difficult to detect. However, if 
the conformal invariance hypothesis is correct, then these non-Gaussian 
correlations should exist at some level, in distinction to the simplest 
inflationary scenarios. Their amplitude is model dependent and possibly much 
larger than the above naive estimate. The present WMAP data does not show 
any evidence of these non-Gaussian statistics \cite{Kom}. Again we are in 
need of a calculation of the amplitude of the non-Gaussianity based on a 
more complete model. In the meantime any detection of non-Gaussian 
statistics of the CMB would be an important clue to their
origin and possibly an important test for the hypothesis of 
conformal invariance.

For higher point correlations, conformal invariance does not 
determine the total angular dependence. Already the four-point 
function takes the form,
\begin{equation}
\langle{\cal O}_{\Delta} (x_1) {\cal O}_{\Delta} (x_2) 
{\cal O}_{\Delta} (x_3) {\cal O}_{\Delta} (x_4)\rangle
\sim { A_4
\over {\prod_{i<j} r_{ij}^{2\Delta/3}} }\ ,
\end{equation}
where the amplitude $A_4$ is an arbitrary function of the two 
cross-ratios, 
$r_{13}^2 r_{24}^2/r_{12}^2 r_{34}^2$ and 
$r_{14}^2 r_{23}^2/r_{12}^2 r_{34}^2$.
Analogous expressions hold for higher $p$-point functions. However in the 
equilateral case $\theta_{ij}=\theta$, the coefficient amplitudes 
$A_p$ become constants and the angular dependence is again completely 
determined. The result is
\begin{equation}
C_p(\theta)\sim (1-\cos\theta)^{{p\over 2}(2-\Delta)}\ ,
\end{equation}
and the expansion in multiple moments yields coefficients $c_{\ell}^{(p)}$ 
of the same form as in Eqn.~(\ref{cl3}) with $3/2$ replaced by $p/2$.
In the limit $\Delta =2$, we obtain the universal 
$\ell$-dependence $\ell(\ell+1)c_{\ell}^{(p)}=6c_2^{(p)}$.

Thus the conformal invariance hypothesis applied to the primordial 
density fluctuations predicts deviations both from the classical 
Harrison-Zel'dovich spectrum and Gaussian statistics, which should be 
imprinted on the CMB anisotropy. A particular realization of this hypothesis 
is provided by the metric fluctuations induced by the known trace anomaly of 
massless matter fields which gives rise to fixed point with a spectral index 
$n > 1$. Although this is disfavored by the WMAP data, lacking a complete 
cosmological model which takes dark energy, dark matter, the CMB and possibly 
other effects into account in a consistent way, it is premature to draw a 
final conclusion on the conformal invariance hypothesis. The
possibility of explaining the small value of $\lambda$ in a natural
way is a strong reason to pursue a more complete cosmological model
within this framework. In order to establish the firm theoretical basis 
for a consistent cosmological model, we return to consideration of the full 
EFT of gravity, reporting on recent progress in the auxiliary field 
description of the non-local terms (\ref{nonl}) generated by the trace anomaly.

\section{The Low Energy EFT of Gravity}

The quantization of the conformal factor in certain specialized cases
carried out in the $1990's$ and reviewed in Secs. 4 and 5 shows that there 
are new degrees of freedom and infrared quantum effects in gravity which are 
not contained in classical general relativity. Moreover these new scalar degrees 
of freedom and their fluctuations appear to be the relevant ones for cosmological
scales and the screening of the cosmological term, which was our initial motivation 
for studying them. In order to place these new degrees of freedom on a solid 
footing together with Einstein's theory, we give in this section a self-contained
systematic effective field theory (EFT) treatment of four dimensional gravity.

The EFT of gravity is determined by the same general principles as 
in other contexts \cite{DGH}, namely by an expansion in powers of derivatives of 
local terms consistent with symmetry. Short distance effects are parameterized 
by the coefficients of local operators in the effective action, with higher order 
terms suppressed by inverse powers of an ultraviolet cutoff scale $M$. The 
effective theory need not be renormalizable, as indeed Einstein's theory is not, 
but is expected nonetheless to be quite insensitive to the details of the 
underlying microscopic degrees of freedom, because of decoupling \cite{DGH}. It 
is the decoupling of short distance degrees of freedom from the macroscopic 
physics that makes EFT techniques so widely applicable, and which we assume 
applies also to gravity.

As a covariant metric theory with a symmetry dictated by the Equivalence Principle, 
general relativity may be regarded as just such a local EFT, truncated at second 
order in derivatives of the metric field $g_{ab}(x)$ \cite{Dono}. When quantum 
matter is considered, the stress tensor $T_a^{\ b}$ becomes an operator. 
Because the stress tensor has mass dimension four, containing up to quartic
divergences, the proper covariant renormalization of this operator requires 
fourth order terms in derivatives of the metric. However the effects of such 
higher derivative {\it local} terms in the gravitational effective action are 
suppressed at distance scales $L \gg L_{Pl}$ in the low energy EFT limit. 
Hence surveying only local curvature terms, it is often tacitly assumed that 
Einstein's theory contains all the low energy macroscopic degrees of freedom 
of gravity, and that general relativity cannot be modified at macroscopic 
distance scales, much greater than $L_{Pl}$, without violating general 
coordinate invariance and/or EFT principles. As we have argued previously
in two dimensions, this presumption should be re-examined in the presence of 
quantum anomalies.

When a classical symmetry is broken by a quantum anomaly, the naive decoupling of 
short and long distance physics assumed by an expansion in local operators with 
ascending inverse powers of $M$ fails. In this situation even the low energy 
symmetries of the effective theory are changed by the presence of the anomaly, 
and some remnant of the ultraviolet physics survives in the low energy description. 
An anomaly can have significant effects in the low energy EFT because it is not 
suppressed by any large energy cutoff scale, surviving even in the limit 
$M \rightarrow \infty$. Any explicit breaking of the symmetry in the classical 
Lagrangian serves only to mask the effects of the anomaly, but in the right 
circumstances the effects of the non-local anomaly may still dominate the local 
terms. We have mentioned in Sec. I the chiral anomaly in QCD with massless quarks, 
whose effects are unsuppressed by any inverse power of the EFT ultraviolet cutoff 
scale, in this case $M \sim \Lambda_{QCD}$. Although the quark masses are non-zero, 
and chiral symmetry is only approximate in nature, the chiral anomaly gives the 
dominant contribution to the low energy decay amplitude of $\pi^0\rightarrow 2\gamma$ 
in the standard model \cite{Adler,BGMF}, a contribution that is missed entirely by a 
local EFT expansion in pion fields. Instead the existence of the chiral 
anomaly requires the explicit addition to the local effective action of a 
{\it non-local} term in four physical dimensions to account for its 
effects \cite{WZ,DGH}.

Although when an anomaly is present, naive decoupling between the short and 
long distance degrees of freedom fails, it does so in a well-defined way, 
with a coefficient that depends only on the quantum numbers of the underlying 
microscopic theory. In fact, since the chiral anomaly depends on the color 
charge assignments of the short distance quark degrees of freedom, the 
measured low energy decay width of $\pi^0\rightarrow 2\gamma$ affords a 
clean, non-trivial test of the underlying microscopic quantum theory of QCD 
with three colors of fractionally charged quarks \cite{DGH,BGMF,Wein}. The 
bridge between short and long distance physics which anomalies provide is 
the basis for the anomaly matching conditions \cite{tHo}. 
 
The low energy effective action for gravity in four dimensions 
contains first of all, the local terms constructed from the
Riemann curvature tensor and its derivatives and contractions up to
and including dimension four. This includes the usual Einstein-Hilbert
action of general relativity,
\begin{equation}
S_{EH}[g] = \frac{1}{16\pi G} \int\, d^4x\,\sqrt{-g}\, (R-2\Lambda)  
\label{clfour}
\end{equation}
as well as the spacetime integrals of the fourth order curvature
invariants,
\begin{equation}
S_{local}^{(4)}[g] = \frac{1}{2} \int\, \sqrt{-g}\, (\alpha C_{abcd}C^{abcd}
+ \beta R^2)\, d^4x\,,
\label{R2}
\end{equation}
with arbitrary dimensionless coefficients $\alpha$ and $\beta$. There are two 
additional fourth order invariants, namely $E= ^*\hskip-.2cm R_{abcd}\,
^*\hskip-.1cm R^{abcd}$ and $\sq R$, which could be added to (\ref{R2}) as well, 
but as they are total derivatives yielding only a surface term and no local 
variation, we omit them. All the possible local terms in the effective action 
may be written as the sum,
\begin{equation}
S_{local}[g] = {1\over 16\pi G}\int\,d^4x\,\sqrt{-g}\,(R-2\Lambda) +
S_{local}^{(4)} + \sum_{n=3}^{\infty} S_{local}^{(2n)}\,.
\label{locsum}
\end{equation}
with the terms in the sum with $n\ge 3$ composed of integrals of local curvature 
invariants with dimension $2n \ge 6$, and suppressed by $M^{-2n + 4}$ at energies 
$E \ll M$. Here $M$ is the ultraviolet cutoff scale of the low energy effective 
theory which we may take to be of order $M_{pl}$. The higher derivative terms 
with $n \ge 3$ are irrelevant operators in the infrared, scaling with negative 
powers under global rescalings of the metric, and may be 
neglected at macroscopic distance scales. On the other hand the two terms in 
the Einstein-Hilbert action $n= 0, 1$ scale positively, and are clearly relevant 
in the infrared. The fourth order terms in (\ref{R2}) are neutral under 
such global rescalings. 

The exact quantum effective action also contains non-local terms in general. 
All possible terms in the effective action (local or not) can be classified 
according to how they respond to global Weyl rescalings of the metric. If the 
non-local terms are non-invariant under global rescalings, then they  
scale either positively or negatively under (\ref{confdef}). If $m^{-1}$ 
is some fixed length scale associated with the non-locality, arising for 
example by the integrating out of fluctuations of fields with mass $m$, 
then at much larger macroscopic distances ($mL \gg 1$) the non-local 
terms in the effective action become approximately local. The terms which 
scale with positive powers of $e^{\sigma_0}$ are constrained by general 
covariance to be of the same form as the $n=0,1$ Einstein-Hilbert terms 
in $S_{local}$, (\ref{clfour}). Terms which scale negatively with 
$e^{\sigma_0}$ become negligibly small as $mL \gg 1$ and are infrared 
irrelevant at macroscopic distances. This is the expected decoupling of 
short distance degrees of freedom in an effective field theory description, 
which are verified in detailed calculations of loops in massive field 
theories in curved space. The only possibility for contributions 
to the effective field theory of gravity at macroscopic distances,
which are not contained in the local expansion of (\ref{locsum}) arise 
from fluctuations not associated with any finite length scale, {\it i.e.} 
$m=0$. These are the non-local contributions to the low energy EFT which 
include those associated with the anomaly. 

The non-local form of the anomalous effective action was given in (\ref{nonl}).
To cast this into local form and exhibit the new scalar degrees of freedom the
$S_{anom}$ contains, it is convenient as in the two dimensional case to 
introduce auxiliary fields. Two scalar auxiliary fields satisfying
\begin{eqnarray}
&& \Delta_4\, \varphi = {1\over 2} \left(E - {2\over 3} \sq R\right)\,,
\nonumber\\
&& \Delta_4\, \psi = {1\over 2} F\,,
\label{auxeom}
\end{eqnarray}
\noindent respectively may be introduced, corresponding to the two non-trivial
cocycles of the $b$ and $b'$ terms in the anomaly \cite{MazMot}. It is then easy 
to see that 
\begin{eqnarray} 
&& S_{anom}[g;\varphi,\psi] =  {b'\over 2}\,\int\,d^4x\,\sqrt{-g}\ 
\left\{ -\varphi \Delta_4 \varphi + \left(E - {2\over 3} \sq R\right) \varphi \right\}
\nonumber\\
&& \quad + {b\over 2} \,\int\,d^4x\,\sqrt{-g}\ \left\{ -\varphi \Delta_4 \psi 
-\psi \Delta_4 \varphi + F \varphi + \left(E - {2\over 3} \sq R\right) 
\psi \right\}
\label{locaux}
\end{eqnarray}
is the desired local form of the anomalous action (\ref{nonl}) \cite{BFS,MotV}. 
Indeed the variation of (\ref{locaux}) with respect to the auxiliary fields
$\varphi$ and $\psi$ yields their Eqs. of motion (\ref{auxeom}), which
may be solved for $\varphi$ and $\psi$ by introducing the Green's function 
$\Delta_4^{-1}(x,x')$. Substituting this formal solution for the auxiliary fields 
into (\ref{locaux}) returns (\ref{nonl}). The local auxiliary field
form (\ref{locaux}) is the most useful and explicitly contains two new scalar 
fields satisfying the massless fourth order wave equations (\ref{auxeom})
with fourth order curvature invariants as sources. The freedom to add homogeneous 
solutions to $\varphi$ and $\psi$ corresponds to the freedom to define
different Green's functions inverses $\Delta_4^{-1}(x,x')$ in (\ref{nonl}).
The auxiliary scalar fields are new local massless degrees of freedom of four 
dimensional gravity, not contained in the Einstein-Hilbert action.

The terms in the classical Einstein-Hilbert action scale with positive 
powers ($L^4$ and $L^2$) under rescaling of distance, and are clearly 
relevant operators of the low energy description. The non-local anomalous terms,
rendered local by the introduction of the auxiliary fields $\varphi$ and $\psi$
scale logarithmically ($\sim\log L$) with distance under Weyl rescalings. Unlike
local higher derivative terms in the effective action, which are either neutral 
or scale with negative powers of $L$, the anomalous terms should not be discarded 
in the low energy, large distance limit. The auxiliary fields are new local scalar 
degrees of freedom of low energy gravity, not contained in classical general 
relativity. The addition of the anomaly term(s) to the low energy effective action 
of gravity amounts to a non-trivial infrared modification of general relativity, 
fully consistent with both quantum theory and the Equivalence Principle \cite{MazMot}. 

The fluctuations generated by $S_{anom}$ define a non-perturbative Gaussian 
infrared fixed point, with conformal field theory anomalous dimensions analogous 
to the two dimensional case \cite{AntM,Odint}. This is possible only because new low 
energy degrees of freedom are contained in $S_{anom}$ which can fluctuate 
independently of the local metric degrees of freedom in $S_{EH}$. Thus the 
effective action of the anomaly $S_{anom}$ should be retained in the EFT of 
low energy gravity, which is specified then by the first two strictly relevant 
local terms of the classical Einstein-Hilbert action (\ref{clfour}), and the 
logarithmic $S_{anom}$, {\it i.e.}
\begin{equation}
S_{eff}[g] = S_{EH}[g] + S_{anom}[g]
\label{Seff}
\end{equation}
contains all the infrared relevant terms in low energy gravity for 
$E \ll M_{pl}$. 

The low energy (Wilson) effective action (\ref{Seff}), in which infrared 
irrelevant terms are systematically neglected in the renormalization
group program of critical phenomena is to be contrasted with the exact 
(field theoretic) effective action, in which the effects 
of all scales are included in principle, at least in the approximation 
in which spacetime can be treated as a continuous manifold. Ordinarily, 
{\it i.e.} absent anomalies, the Wilson effective action should contain 
only {\it local} infrared relevant terms consistent with symmetry \cite{RG}. 
However, like the anomalous effective action generated by the chiral 
anomaly in QCD, the non-local $S_{anom}$ must be included in the low 
energy EFT to account for the anomalous Ward identities, even in the 
zero momentum limit, and indeed logarithmic scaling with distance 
indicates that $S_{anom}$ is an infrared relevant term. Also even if 
no massless matter fields are assumed, the quantum fluctuations of 
the metric itself will generate a term of the same form as $S_{anom}$
\cite{AMMc}. 

By using the definition of $\Delta_4$ and integrating by parts, 
we may express the anomalous action also in the form,
\begin{equation} 
S_{anom} = b' S^{(E)}_{anom} + b S^{(F)}_{anom}\,,
\label{allanom}
\end{equation}
with
\begin{eqnarray}
&&\hskip-2.5cm S^{(E)}_{anom} \equiv {1\over 2} \int d^4x\sqrt{-g}\left\{
-\left(\sq \varphi\right)^2 + 2\left(R^{ab} - {R\over 3}g^{ab}\right)(\nabla_a \varphi)
(\nabla_b \varphi) + \left(E - {2\over 3} \sq R\right) \varphi\right\},\nonumber\\
&& \hskip-2.5cm S^{(F)}_{anom} \equiv \int d^4x \sqrt{-g}
\left\{ -\left(\sq \varphi\right)
\left(\sq \psi\right) + 2\left(R^{ab} - {R\over 3}g^{ab}\right)(\nabla_a \varphi)
(\nabla_b \psi)\right.\nonumber\\
&& \quad\quad + \left.{1\over 2} F \varphi + 
{1\over 2} \left(E - {2\over 3} \sq R\right) \psi \right\}
\label{SEF}
\end{eqnarray}
It is this final local auxiliary field form of the effective action
which is to be added to classical Einstein-Hilbert action to obtain
the effective action of low energy gravity in (\ref{Seff}).
We note also that in this form the simple shift of the auxiliary 
field $\varphi$ by a spacetime constant, 
\begin{equation}
\varphi \rightarrow \varphi + 2\sigma_0
\end{equation}
yields the entire dependence of $S_{anom}$ on the global Weyl rescalings
(\ref{confdef}), {\it viz.}
\begin{eqnarray}
&& S_{anom}[g; \varphi, \psi] \rightarrow 
S_{anom}[e^{2\sigma_0} g; \varphi + 2 \sigma_0, \psi] \nonumber\\
&& \qquad = S_{anom}[g; \varphi, \psi] + \sigma_0
\,\int\,d^4x\,\sqrt{-g}\, \left[ bF + b'
\left(E - {2\over 3} \sq R\right)\right]\,,
\end{eqnarray}
owing to the strict invariance of the terms quadratic in the auxiliary
fields under (\ref{confdef}) and eqs. (\ref{FEsig}). Thus the auxiliary field
form of the anomalous action (\ref{SEF}) contains the same information
about the global Weyl anomaly and large distance scaling as $\Gamma_{WZ}$.  

One may ask if there are any other modifications of classical general
relativity at low energies that are consistent with general covariance
and EFT principles. The complete classification of the terms in 
the exact effective action \cite{MazMot,MotV} into just three categories means 
that all possible infrared relevant terms in the low energy EFT, which are not 
contained in $S_{local}$ of (\ref{locsum}) must fall into $S_{anom}$, {\it i.e.} 
they must correspond to non-trivial co-cycles of the local Weyl group. The Weyl 
invariant terms in the exact effective action are by definition
insensitive to rescaling of the metric at large distances. Hence these 
(generally quite non-local) terms do not give rise to infrared relevant terms 
in the Wilson effective action for low energy gravity.

Furthermore, the form of the non-trivial co-cycles in $S_{anom}$ is severely 
restricted by the locality and general covariance of quantum field theory. The 
ultraviolet divergences in the stress-energy tensor of quantum fields in curved 
spacetime are purely local. It is these divergences when renormalized 
consistently with covariant conservation of the local operator $T_{ab}(x)$ that 
give rise to the purely local form of the trace anomaly. Since all the local 
gauge invariant terms with mass dimension four matched to the physical dimension 
of spacetime are easily cataloged, the only non-trivial terms in $S_{anom}$ at 
low energies which can arise from short distance renormalization effects are 
exactly those generated by the known form of the local trace anomaly 
(\ref{tranom}). Decoupling fails in local quantum field theory only in the 
very narrow and well-defined way dictated by local anomalies, and these 
uniquely determine the non-local additions to the effective action, up to 
any contributions from $S_{inv}$, which in any case have negligibly small 
effect on very large distance scales. The form of the effective action $S_{anom}$ 
at macroscopic distances $L \gg L_{pl}$ is not expected to change substantially 
even if the condition of strict locality of the underlying quantum theory is 
relaxed or replaced eventually by a more fundamental, microscopic description 
of gravitational interactions at very large mass scales of order $M_{pl}$. If 
this were not the case, then the classical Einstein theory could be overwhelmed 
by all sorts of non-local quantum corrections from unknown microscopic physics, 
and would lose all predictive power for macroscopic gravitational phenomena. 
Instead, under the defining assumptions of general covariance and the EFT 
hypothesis of decoupling of physics associated with massive degrees of freedom, 
the infrared modification of Einstein's theory specified by (\ref{Seff})-(\ref{SEF}) 
is tightly constrained and becomes essentially unique. Notice in particular that 
the EFT logic precludes any inverse powers of the Ricci scalar or other local 
curvature invariants appearing in the denominators of terms in the effective 
action. It is the EFT of gravity defined by (\ref{Seff}) which is the basis for 
the new degrees of freedom and their fluctuations leading to the conformal fixed 
point of Secs 4 and 5. It should also be the basis upon which a more comprehensive 
cosmological model of dark energy incorporating these effects is constructed.

\section{The Horizon Boundary}

Quantum effects in global de Sitter space indicate infrared effects on the
horizon scale are important. These effects are contained in two new infrared
relevant operators in the EFT of low energy gravity described by (\ref{allanom}) 
above. If we hope to construct a consistent model of dark energy in which the
quantum effects of the trace anomaly can modify classical theory, then we must
identify first where the effects of the new terms in the EFT can be significant. 
This involves again a careful investigation of horizon and near horizon effects 
in de Sitter spacetime.

Two familiar examples of spacetimes with horizons are the Schwarzschild metric 
of an uncharged non-rotating black hole, and the de Sitter metric. Both can be 
expressed in static, spherically symmetric coordinates in the form,
\begin{equation}
ds^2 = -f(r)\, c^2 dt^2 + {dr^2 \over f(r)} + 
r^2\left( d\theta^2 + \sin^2\theta\,d\phi^2\right)
\,.
\label{sphsta}
\end{equation}
In the Schwarzchild case, 
\begin{equation}
f_{_S}(r) = 1 - \frac{r_{_S}}{r}\,,\qquad r_{_S} = {2GM\over c^2}\,,
\label{Sch}
\end{equation}
while in the de Sitter case, from (\ref{dSstatic}),
\begin{equation}
f_{_{dS}}(r) = 1 - \frac{r^2}{r_{_H}^2}\,, \qquad r_{_H} = 
\left({3\over \Lambda}\right)^{1\over 2}\,,
\end{equation}
respectively. The first metric approaches flat space at large $r$ but becomes 
singular at the finite radius $r_{_S}$. The second is the static coordinates
of de Sitter spacetime (\ref{dSstatic}), describing the interior of a 
spherical region with a coordinate singularity at the finite radius $r_{_H}$. In 
both cases the metric singularities may be regarded as pure coordinate artifacts, 
in the sense that they can be removed entirely by making a {\it singular} 
coordinate transformation to a different set of well-behaved coordinates in 
the vicinity of the horizons. Indeed by undoing (\ref{ctrans}) we can transform 
de Sitter spacetime back to FLRW coordinates. However, as we have observed already 
in Sec. 2 certain global effects such as the tempearture associated with the 
horizon in the standard treatments cannot be transformed away by a local coordinate transformation. 

It is important to recognize that the Equivalence Principle implies invariance 
under regular coordinate transformations, whereas {\it singular} transformations 
and the analytic extensions of spacetime they involve require a physical assumption, 
namely that there are no stress-energy sources or discontinuities of any kind 
at coordinate singularity. Even in the classical theory the hyperbolic nature of 
Einstein's equations allows for sources and/or discontinuities transmitted at 
the speed of light on a null hypersurface, such as the Schwarzschild or
de Sitter horizon. Analytic continuation amounts then to a physical
{\it assumption} that no such discontinuities are present.

Moreover, when quantum fields are considered, matter is no longer described 
as pointlike particles following classical geodesics, but as matter waves, 
especially on the horizon scale. This is the origin of the particle
creation and Hawking effects discussed in Sec. 2. The matter wave equations 
such as the Dirac or Klein-Gordon eqs. couple to the electromagnetic and 
metric {\it potentials}, not the local Maxwell or Riemann curvature tensors. 
Hence quantum matter effects can depend on gauge invariant global functions 
of the potentials such as $\exp (i \oint A_\mu dx^\mu)$, through the boundary 
conditions imposed on the solutions. Like the  Aharonov-Bohm effect in 
electron scattering or the Abrikosov vortex in superconductors, a {\it singular} 
``gauge" transformation may have global physical effects at the quantum level, 
even though the field curvature tensor is small or even vanishing nearly 
everywhere. This is because it is not truly an allowed symmetry transformation
of the quantum theory at all. Clearly such global quantum effects of matter waves 
cannot be captured by a description of matter as pointlike particles of 
infinitesimal size following classical geodesics.

These quantum wavelike effects show up in the behavior of the renormalized
expectation value of the stress-energy tensor, 
$\langle\Psi\vert T_a^{\ b}\vert\Psi\rangle$ as $r \rightarrow r_{_S}$ or 
$r \rightarrow r_{_H}$. This expectation value depends upon the 
quantum state $\vert \Psi\rangle$ of the field theory, specified by choosing 
particular solutions of the wave equation $\sq \Phi = 0$. The Schwarzschild 
or de Sitter horizon is a characteristic surface and regular singular point of 
the wave equation in static coordinates (\ref{sphsta}). Hence the 
general solution of the wave equation is singular there, and so is the 
expectation value, $\langle \Psi\vert T_a^{\ b}\vert\Psi \rangle$ in the 
corresponding quantum state. This generic singular behavior is essentially 
kinematical in origin, since a photon with frequency $\omega$ and energy 
$\hbar \omega$ far from the horizon has a local energy 
$E_{loc}=\hbar \omega f^{-{1\over 2}}$. The stress-energy is dimension 
four, and its generic behavior near the event horizon is $E_{\rm loc}^4 \sim f^{-2}$. 
Calculations of $\langle \Psi\vert T_a^{\ b}\vert\Psi\rangle$ both directly from 
quantum field theory and through the anomaly action (\ref{locaux}) show this 
diverging behavior in the vacuum state defined by absence of quanta with respect 
to the static time coordinate $t$ in (\ref{sphsta}) \cite{MotV, Boul,ChrFul}. In 
this ``vacuum" state the stress tensor behaves like the {\it negative} of a 
radiation fluid at the local temperature, 
$T_{loc} = \hbar c\vert f'\vert f^{-{1\over 2}}/4\pi k_{_B}$. Hence 
the ``vacuum" near a spacetime horizon is sensitive to arbitarily high
frequency components of quantum fluctuations, whose backreaction effects
through the stress tensor expectation, $\langle T_a^{\ b}\rangle$ may
become arbitrarily large.

Although this has been known for some time \cite{Boul,ChrFul}, the attitude usually 
adopted is that states which lead to such divergences on the horizon are to be 
excluded, and only states regular on the horizon are allowed. However this 
particular boundary condition is only one possibility and is not required by 
any general principles of quantum field theory in one causally connected region 
of spacetime. The essence of an event horizon is that it divides spacetime into 
regions which are causally disconnected from each other. In both the de Sitter 
and Schwarzschild cases, there are globally regular states such as the 
Bunch-Davies and Hartle-Hawking states respectively \cite{BD,HarHaw}, but these 
specify that precise quantum correlations be set up and maintained in regions of 
the globally extended spacetimes which never have been in any causal contact 
with each other. Despite their mathematical appeal, it is by no means clear 
physically why one should restrict attention to quantum states in which exact 
phase correlations between causally disconnected regions are to be rigorously 
enforced. This is the quantum version of the causality or horizon problem 
encountered with respect to the CMB in classical FLRW cosmologies. As soon as 
one drops this acausal requirement, and on the contrary restricts attention to 
states with correlations that could have been arranged causally in the past, 
within the region $r \le r_{_H}$ in de Sitter spacetime, for example, then 
states with divergent $\langle \Psi\vert T_a^{\ b}\vert\Psi \rangle$ on the 
horizon become perfectly admissible (ignoring for the moment the large 
backreaction such a stress-energy must exert on the background geometry). This 
is the conclusion reached also by consideration of the solutions of the auxiliary 
field equations (\ref{auxeom}) in Schwarzschild and de Sitter spacetimes \cite{MotV}. In fact, the $\varphi$ 
and $\psi$ fields generally {\it diverge} on the horizon, even though the 
local Riemann curvature is classically small there. The quantum EFT which 
allows these states is then quite different in its basis and consequences 
from an arbitrary presciption to exclude them {\it a priori}.

Again our experience with the Casimir effect suggests a physical interpretation
and resolution of the divergences in these states. In the calculation of the local 
stress tensor $\langle \Psi\vert T_a^{\ b}\vert\Psi \rangle$ in flat spacetime 
with boundaries, one also finds divergences in the generic situation of 
non-conformally invariant fields and/or curved boundaries \cite{CanDeu}. The 
divergences may be traced to the boundary conditions imposed on modes of the high 
frequency components of a quantum field, and cannot be removed by standard 
renormalization counterterms in the bulk. As the theory and applications of the 
Casimir effect have developed, it became clear that the material properties of 
the real conductors involved in the experiments must be taken into account
to regulate these mathematical divergences. Casimir's idealized boundary conditions 
on the electromagnetic field, appropriate for a perfect conductor with infinite 
conductivity, must give way in more realistic calculations to boundary conditions 
incorporating the finite conductivity response function of real metals \cite{Moh}. 
The idealized boundary conditions which led to the divergences are not to be excluded 
from consideration by mathematical {\it fiat}; indeed they are the correct boundary 
conditions at low to moderate frequencies, and the local stress tensor would continue
to grow larger as the boundary is approached, if no new physics were to intervene.
The appropriate modification of the boundary conditions at higher frequencies and
the cutoff of this growth is obtained by correctly incorporating the physics of the 
conducting boundary. Then finite results confirmed in detail by experiment are 
obtained \cite{Moh}. At still smaller length and time scales approaching atomic 
dimensions, the approximation of a continuous or average conductivity response 
function of the metal surface will have to be modified again, to take account the 
electron band structure and microscopic graininess of the conductors, which are 
composed finally of discrete atomic constituents. Only the physical response function 
of the metal, not the mathematics of analytic continuation (which involves an unchecked
assumption of arbitrarily high Fourier components) can determine the correct boundary
conditions to be imposed at the boundary, and the behavior of the stress tensor as 
the boundary is approached.

In the case of the Schwarzschild or de Sitter horizon this boundary condition
requires additional physical input. The wave equation in the spherically 
symmetric static coordinates (\ref{sphsta}) can be separated by writing 
$\Phi = e^{-i\omega t} Y_{lm} {\psi_{\omega l}\over r}$, and
the second order ordinary differential equation for the radial function,
$\psi_{\omega l}$ may be cast in the form,
\begin{equation}
\left[-{d^2\over dr^{*2}} + V_l\right] \psi_{\omega l} = \omega^2 \psi_{\omega l}\,.
\label{req}
\end{equation}
The change of radial variable from $r$ to $r^* \equiv \int^r\,\frac{dr} {f(r)}$
has been made in order to put the second derivative term into standard form, and the
scattering potential for the mode with angular quantum number $l$ is 
\begin{equation}
V_l = f \left[ \frac{1}{r} \frac{df}{dr} + \frac{l(l+1)}{r^2} \right]\,.
\label{pot}
\end{equation}
Since $f\rightarrow 0$ linearly as $r$ approaches the horizon, the variable
$r^*\rightarrow -\infty$ logarithmically, and the potential goes to zero there, 
vanishing exponentially in $r^*$. Hence the solutions of (\ref{req}) define 
one dimensional scattering states on an infinite interval (in $r^*$), and the 
boundary conditions on the horizon that ensure that the scattering matrix is 
hermitian are {\it free} ones, allowing both incoming and outgoing plane wave 
modes as $r^*\rightarrow -\infty$. The vacuum state defined by zero occupation
number with respect to these scattering states is the Boulware vacuum $\vert \Psi_B\rangle$, which has a divergent $\langle \Psi_B\vert T_a^{\ b}\vert\Psi_B \rangle$, behaving
in fact like $-T_{\rm loc}^4\ {\rm diag}(-3,1,1,1) \sim f^{-2}$ on the horizon 
\cite{Boul,ChrFul}. In contrast to the Casimir effect in flat spacetime with curved 
boundaries, this divergence does not arise from hard Dirichlet or Neumann boundary 
conditions, but from an infinite redshift surface with free boundary conditions. 
Hence the properties of no ordinary material at the boundary can remove this divergence, 
and the effective cutoff of horizon divergences can arise only from new physics in the 
gravitational sector at ultrashort scales, {\it i.e.} in the structure of spacetime 
itself very near to the horizon.

The divergence of the expectation value $\langle T_a^{\ b}\rangle$ at the horizon
is completely described by the auxiliary potentials of the anomalous action
(\ref{allanom})-(\ref{SEF}) \cite{MotV}. As expected from their derivation and the 
analogous situation in two dimensions, these auxiliary fields carry non-local 
information about the global quantum state and boundary conditions. Their
fluctuations describe the higher point correlators of the stress-energy tensor 
in the given quantum state. From the particle creation, thermodynamic and 
fluctuation-dissipation discussion in Sec. 2, and the conformal fixed point 
considerations of Secs. 4 and 5, it is these fluctuations and the higher point 
correlators of the quantum $T_a^{\ b}$ that generate the backreaction on the mean 
geometry necessary to relax the effective cosmological term to zero. These
become significant in the limit $f(r) \rightarrow 0$ as the horizon boundary
is approached. Thus the locus of important quantum effects from $S_{anom}$ is 
not on superhorizon scales in the FLRW coordinates (\ref{RWflat}), but in a 
very thin boundary layer very close to the Schwarzschild or de Sitter horizon 
in static coordinates. In other words, the physical location of the conformally 
invariant phase of gravity discussed  somewhat abstractly in Secs. 4 and 5 should 
be just in this boundary layer very close to $r= r_{_H}$. 

\section{A New Cosmological Model of Dark Energy}

The suggestion that a quantum phase transition may occur in the vicinity of the 
classical Schwarzschild horizon $r_{_S}$ has been made in \cite{CHLS} and 
\cite{star}. The fluctuations of the auxiliary fields of $S_{anom}$ which we 
found previously describe a conformally invariant phase of gravity with vanishing 
cosmological term, and may be responsible for this transition near the horizon. 
At a first order phase transition in which the quantum ground state rearranges 
itself, the vacuum energy of the state can change. Hence the region interior 
to $r_{_S}$ of the Schwarzschild geometry may have a different effective value 
of $\Lambda$ than the exterior region. We have suggested that the cosmological 
term itself may be viewed as the order parameter of a kind of gravitational 
Bose-Einstein condensate (GBEC), and the phase transition near the horizon where 
this condensate disorders would then become similar to BEC phase transition 
observed in cold atomic systems \cite{star}. Moreover, since due to (\ref{accond}) 
the vacuum dark energy equation of state with $\rho_v = - p_v > 0$ acts as an 
effective repulsive term in Einstein's equations, a positive value of $\Lambda$ 
in the interior serves to support the system against further gravitational collapse. 
For this to work the effective value of $\Lambda$ in the interior would have to 
adjust itself dynamically to the total mass of the system in order to reach a 
non-singular state of stable equilibrium with $r_{_H} \simeq r_{_S}$. The 
de Sitter interior is free of any singularities and the entropy of this state 
is much less than the Bekenstein-Hawking entropy of a black hole. It therefore 
suffers from no ``information paradox" \cite{star}. 

It is interesting to remark that the non-singular configuration described in 
\cite{star} may be viewed as the gravitational analog of the model of an 
electron, which was one of the motivations of some of the original 
investigations of the Casimir effect. Since the Casimir force on 
a conducting charged sphere is {\it repulsive}, it cannot cancel the 
classical repulsive Coulomb self-force \cite{Boy}. However a repulsive Casimir 
force with interior vacuum energy $p_v = - \rho_v < 0$ is exactly what is needed 
to balance the {\it attractive} force of gravity to prevent collapse to a singularity. 
The Casimir proposal to model an elementary particle such as the electron as a 
conducting spherical shell does not work as originally proposed, but the analogous 
model for the non-singular final state of gravitational collapse of a macroscopic 
self-gravitating object \cite{Dir} appears to be perfectly viable. 

The model we arrive at is one with the de Sitter interior matched to a Schwarzschild
exterior, sandwiching a thin shell which straddles the region near to 
$r_{_H} \simeq r_{_S}$, cutting off the divergences in $\langle T_a^{\ b} \rangle$ 
as $r_{_H}$ is approached from inside and $r_{_S}$ is approached from outside. This 
thin shell is the boundary layer where the new physics of a quantum phase transition 
takes place. In the EFT approach this new physics is described by the fluctuations 
of the auxiliary scalar degrees of freedom $\varphi$ and $\psi$ in $S_{anom}$. 

In a true quantum boundary layer, fluctuations in all the higher point correlators
of $T_a^{\ b}$ are to be expected. This boundary layer is therefore quite
non-perturbative. In effect the coupling constant $\lambda$ is multiplied by
inverse factors of $f(r)$ which greatly enhance the quantum effects in the
boundary layer. Thus even if $\lambda \ll 1$ a critical surface is reached
when one approaches the horizon boundary from the interior de Sitter phase.
As a first approximation we may treat the quantum boundary layer 
in a mean field approximation, in which Einstein's equations continue to hold, 
but with an effective equation of state of the material making up the layer.
This ``material" is the quantum vacuum itself, with a stress tensor described
by the auxiliary scalar fields of the effective action (\ref{allanom})-(\ref{SEF}).
Then the divergences in this stress tensor are cut off by backreaction on the 
classical geometry, replacing an infinite redshift surface at the horizon with 
a finite one. This ultrarelativistic vacuum effect at a causal boundary suggests 
that the most extreme equation of state consistent with causality should play a 
role here, namely the Zel'dovich equation of state $p=\rho$, where the speed of 
sound becomes equal to the speed of light. This is the critical equation of
state at the limit of stability for a phase transition to a new phase with a 
different value of the vacuum energy. It also arises naturally as one component 
of the stress-energy tensor, $\langle T_a^{\ b}\rangle = \ {\rm diag} 
(-\rho, p, p_{\perp}, p_{\perp})$ in a state such as the Boulware state.
The conservation equation,
\begin{equation}
\nabla_a T^a_{\ r} = {d p\over dr} + {\rho + p\over 2f} \,{d f\over dr} 
+ 2\ \frac{ p-p_{\perp}}{r}= 0\,,
\label{cons}
\end{equation}
implies three independent components in the most general static, spherically 
symmetric case. It is clear from (\ref{cons}) that the three independent components 
can be taken to be that with $p = \rho/3$, behaving like $f^{-2}$, $p=\rho$, 
behaving like $f^{-1}$, and $p=-\rho$, behaving like $f^{0}$, reflecting the 
allowed dominant and subdominant classical scaling behaviors of the stress 
tensor near the horizon.

In the simplest model possible we make the further approximation of setting 
the tangential pressure $p_{\perp} = p$ and consider only two independent 
components of the stress tensor in non-overlapping regions of space. In that 
case we have three regions, namely,
\begin{equation}
\begin{array}{clcl}
{\rm I.}\ & {\rm Interior\ (de Sitter):}\ & 0 \le r < r_1\,,\ &\rho = - p \,,\\
{\rm II.}\ & {\rm Thin\ Shell:}\ & r_1 < r < r_2\,,\ &\rho = + p\,,\\
{\rm III.}\ & {\rm Exterior\ (Schwarzschild):}\ & r_2 < r\,,\ &\rho = p = 0\,.
\end{array}
\end{equation}
Because of (\ref{cons}), $p=-\rho$ is a constant in the interior, which
becomes a patch of de Sitter space in the static coordinates (\ref{sphsta}), for
$0\le r \le r_1 < r_{_H}$. The exterior region is a patch of Schwarzschild spacetime
for $r_{_S} < r_2 \le r < \infty$. The $p= \rho/3$ component of the stress tensor
and the smooth transition that it would make possible from one region to another has 
been neglected in this simplest model. In the Boulware state this $p= \rho/3$
traceless stress tensor has negative sign near $r_{_H}$ or $r_{_S}$. Tangential
stresses have been considered by the authors of \cite{Vis}.

The location of the interfaces at $r_1$ and $r_2$ can be estimated by the behavior
of the stress tensor near the Schwarzschild and de Sitter horizons. If 
$1 - r_{_S}/r_1$ is a small parameter $\epsilon$, then the location of the 
outer interface occurs at an $r_1$ where the most divergent term in the local 
stress-energy $\propto M^{-4}\epsilon^{-2}$, becomes large enough to affect the 
classical curvature $\sim M^{-2}$, {\it i.e.} for
\begin{equation}
\epsilon \sim \frac{M_{\rm pl}} {M} \simeq 10^{-38}\ \left(\frac {M_{\odot}}{M}\right)\,,
\label{epsest}
\end{equation}
where $M_{\rm pl}$ is the Planck mass $\sqrt{\hbar c/G} \simeq 2 \times 10^{-5}$ gm.
Thus $\epsilon \ll 1$ for an object of the order of a solar mass, $M=M_{\odot}$,
with a Schwarzschild radius of order of a few kilometers. 

If instead of a collapsed star one considers the interior de Sitter region to be 
a model of cosmological dark energy, then the radius $r_{_H}$ is set by 
measured value of (\ref{cosmeas}), 
\begin{equation}
r_{_H} = \sqrt{\frac{3}{\Lambda_{meas}}} \simeq 1.5 \times 10^{28}\ {\rm cm}, 
\label{rHcos}
\end{equation}
{\it i.e.} the size of the entire visible universe, and $M \approx 5 \times 10^{22} 
M_{\odot} \simeq 10^{56}$ gm becomes of the order of the total mass-energy 
of the visible universe. In that case $\epsilon \simeq 2 \times 10^{-61}
\simeq \sqrt\lambda$ is very small indeed.

Since the function $f(r)$ is of order $\epsilon \ll 1$ in the transition region II, 
the proper thickness of the shell is 
\begin{equation}
\ell = \int_{r_1}^{r_2}\, dr\,f^{-{1\over 2}}  \sim \epsilon^{1\over 2} r_{_S}
\sim \sqrt {L_{\rm pl} r_{_H}} \ll r_{_H}\,.
\label{thickness}
\end{equation}
Although very small, the thickness of the shell is very much larger than the Planck 
scale (\ref{Lpl}). For $r_{_H}$ given by (\ref{rHcos}), the physical thickness of
the shell is macroscopic: $\ell \approx .04$ mm. The energy density and pressure 
in the shell are of order $M^{-2}$ and far below Planckian for $M\gg M_{pl}$, so that 
the geometry can be described reliably by Einstein's equations, essentially everywhere,
except within the thin shell. The details of the solution in region II, the matching 
at the interfaces, $r_1$ and $r_2$, and analysis of the thermodynamic stability of 
the gravitational vacuum condensate star (`gravastar') were studied in \cite{star}.

We note from (\ref{ctrans}) that in this kind of cosmological model,
the past boundary at $r=r_{_H}$ is at the infinite past of the RW coordinates.
Thus one trades a possible special origin of time and the spacelike
singularity of the big bang in FLRW cosmologies for a special spatial
origin and location of the boundary wall. The redshift of primordial 
radiation is then the gravitational redshift due to the potential change 
from the cosmological horizon to an observer in the interior de Sitter
geometry. For most of the interior volume the effects of the past boundary 
are observed only as relic CMB radiation and a residual vacuum energy, which 
would be difficult to distinguish from a FLRW model far from the boundary. 
Since the universe is apparently $74$\% vacuum dark energy, the
cosmological model first proposed by de Sitter \cite{DES}, in which
$\Omega_{\Lambda}$ is unity becomes again a good first approximation
to the observations. 

This simple model is clearly very far from complete. The boundary layer has been 
posited from the allowed behavior of the stress-energy near the horizon, 
rather than a full solution of the EFT equations for the auxiliary fields 
following from (\ref{Seff}). The value of $\Lambda$ in the interior is 
constant and can take on any value, but the solution has $\Omega_{\Lambda} = 1$ 
with no matter or radiation whatsoever in the interior. Since the fluctuations 
of the auxiliary scalar fields in the trace anomaly action are necessary for 
the dynamical relaxation of the vacuum energy, we can obtain only a solution 
which is static without considering those fluctuations in detail. This static 
solution does not describe the evolution of the dark energy towards smaller 
values with time or the very small present value of $\lambda$. No attempt 
has been made to construct a fully dynamical cosmological model which would 
have to pass the many successful tests of the standard FLRW models, including 
the magnitude, spectrum, and statistics of the CMB. On the other hand if we start 
with a simplified cosmological model of pure dark energy in which $\Omega_{\Lambda}$ is 
exactly one, our challenge becomes to explain why it is actually $0.74$, rather than
unity, instead of misestimating $\lambda$ by $122$ orders of magnitude.
 
Despite its drastic simplifications the interior de Sitter gravastar model 
of dark energy does illustrate the possibility of $\rho_v$ being a kind of 
order parameter which is spacetime dependent, whose value depends on boundary 
conditions at macroscopic scales. Thus the dark energy becomes a boundary
effect, analogous to the Casimir energy (\ref{Casimir}), with the role of the 
conducting plates being taken by a critical boundary layer in the spacetime
vicinity of the cosmological horizon. In order for the bulk vacuum energy
to scale quadratically with $H$, {\it i.e.} $\rho_v \sim c^4/G r_{_H}^2 $,
rather than $\hbar c/r_{_H}^4$ as might be expected from (\ref{Casimir}),
it is necessary for fluctuations of the metric itself, {\it i.e.} the 
stress-energy of gravitational waves to play an important role. These
fluctuations are generated at the horizon boundary near $r_{_H}$. Because
of the anomalous trace terms in (\ref{tranom}), the stress-energy of the 
gravitational waves is not tracefree, but can contain also a trace part 
with $p = - \rho$ of either sign which can change the effective value of 
$\Lambda$ in the interior. Once $\Lambda$ becomes dynamical then its value 
is determined not by naturalness considerations at the UV cutoff scale as 
in (\ref{zeropt}), but by the physics at the boundary $r= r_{_H}$ and its 
dynamical evolution. 

The basic assumption required for a solution of this kind to exist is that 
gravity, {\it i.e.} spacetime itself, must undergo a quantum vacuum 
rearrangement phase transition in the vicinity of the horizon, 
$r\simeq r_{_H}$. Clearly this cannot occur in the strictly classical 
Einstein theory of general relativity with $\Lambda$ constant. It requires 
that the fluctuations $\langle T_a^{\ b}(x) T_c^{\ d} (x')\rangle$ and higher 
stress tensor correlators about its mean value be taken into account near the 
horizon. These higher order correlators are generated in the EFT approach by 
the additional scalar degrees of freedom in $S_{anom}$ given by (\ref{allanom}) 
and (\ref{SEF}). The addition of $S_{anom}$ is the minimal modification of 
Einstein's theory {\it required} by quantum theory and stress tensor 
renormalization consistent with the Equivalence Principle. The new
degrees of freedom allow the vacuum energy density $\rho_v$ to change
and adjust itself dynamically in the interior spacetime. Einstein's
theory otherwise continues to apply almost everywhere, sufficiently far
from the quantum boundary layer, with a dynamical bulk $\rho_v$ coupled to
and determined by the fluctuations at the horizon.

Since allowing the effective value of $\Lambda$ to change with time would require 
the generation of gravitational and other radiation at the horizon boundary,
this would realize the dissipative mechanism of relaxation of coherent vacuum 
energy into matter/radiation modes which we discussed in Sec. 2. From (\ref{ESdS}) 
a continuously decreasing $\rho_v$ necessary to give a cosmologically acceptable 
solution to the problem of dark energy requires a continuous energy inflow through
the cosmological horizon. The stress tensor of such an inflow would be expected 
to contain dissipative terms arising from the bulk viscosity of fluctuations
of $T_a^{\ a}$ from the trace anomaly terms, consistent with the general
fluctuation-dissipation considerations of Sec. 2. This relaxation of $\Lambda$ 
to smaller and smaller values would not require any detailed information about 
Planck scale physics, but instead be consistent with the general hypothesis of 
decoupling, with the only short distance effects essential for macroscopic cosmology coupling at the horizon boundary, described by the effective action (\ref{Seff}). 
In the limit $\lambda \rightarrow 0$, the boundary layer is removed to infinity, 
the auxiliary fields in (\ref{auxeom}) have zero sources, and empty flat space 
is the only stable asymptotic solution.

The analogy with atomic Bose-Einstein condensates may also be a fruitful
one to pursue. The vacuum energy density $\rho_v$ is a kind of gravitational
vacuum condensate \cite{star}. This condensate is self-trapped by its own gravity.
The interior vacuum energy density depends on the total number of ``atoms" in 
the trap. The condensate is described by EFT methods analogous to the 
Gross-Pitaevskii EFT of BEC's in terms of the long wavelength collective 
modes of the system. Near the horizon boundary the condensate disorders, 
due to the quantum fluctuations of the auxiliary degrees of freedom in 
the EFT of gravity. At a finer level of resolution, and in particular 
near the boundary, the continuum mean field description must give way to 
a more fundamental treatment in terms of the analogs of the atomistic degrees 
of freedom that make up the vacuum condensate \cite{Maz}. An important 
step towards such a description would be to include fluctuations about 
the mean field, and their associated dissipative effects. 

Clearly much more work remains to be done before a consistent dynamical theory
of dark energy based on this interconnected set of ideas can be proposed. Yet 
the essential physical basis and EFT elements of such a dynamical theory would 
seem to be in place. Only when a comprehensive cosmological model incorporating 
these effects is available can we determine if it passes all observational tests 
of standard cosmology, and make unambiguous predictions for future measurements.
The prediction of the magnitude of deviations from the classical 
Harrison-Zel'dovich spectrum at large angles and non-Gaussian correlations 
in the CMB remain the most promising tests of the conformal invariance 
hypothesis \cite{sky}. If cosmological dark energy is a finite size effect of 
the universe in the large, whose value is determined by conformal
fluctuations at the infrared horizon scale, its dynamical relaxation to 
smaller values over time provides a natural resolution to the dilemma 
of quantum zero-point energy, originally raised by Pauli eighty years 
ago, but now made urgent by its detection (\ref{cosmeas})-(\ref{OLam}) 
in the cosmos.

\section*{Acknowledgements}
\vskip -.2cm
I. A. was supported in part by the European Commission under the RTN contract MRTN-CT-2004-503369.


\section*{References}
\begin{thebibliography}{10}
\vskip -.2cm

\bibitem{Cas} Casimir H B G 1948 {\it Proc. Kon. Ned. Acad. Wetenschap.} 
{\bf 51} 793; \hfill\break
Sparnaay M J 1958 {\it Physica} {\bf 24} 751.

\bibitem{Moh} Bordag M, Mohideen U and Mostepanenko V M 2001
{\it Phys. Rept.} {\bf 353}, 1 and references therein.

\bibitem{Pau} Pauli W (unpublished); Enz C P and Thellung A 
1960 {\it Helv. Phys. Acta} {\bf 33} 839; \hfil\break
Rugh S E and Zinkernagel H 2002 {\it Stud. Hist. Philos. Mod. Phys.} 
{\bf 33}, 663; \hfil\break
Straumann N 2002 e-print arxiv: gr-qc/0208027. 

\bibitem{Ner} Nernst W 1916 {\it Verh. Dtsch. Phys. Ges.} {\bf 18} 83
had given an estimate of zero point energy similar to (\ref{zeropt}) in 
connection with aether theories even earlier.

\bibitem{SNI} Riess A G {\it et. al.} 1998 {\it Astron. J.} {\bf 116}, 1009; 
2004 {\it Astron. J.} {\bf 607} 665;\\
Perlmutter S {\it et. al.} 1999 {\it Astrophys. J.} {\bf 517} 565;\\
Tonry J L {\it et. al.} 2003 {\it Astrophys. J.} {\bf 594}, 1.

\bibitem{BiD}
Birrell N D and Davies P C W 1982 {\it Quantum Fields in Curved
Space} (Cambridge: Cambridge University Press, Cambridge), and
references therein.

\bibitem{anom} Capper D and Duff M 1974 {\it Nuovo Cimento A} {\bf 23} 173;
1975 {\it Phys. Lett. A} {\bf 53} 361; \hfil\break
Deser S, Duff M and Isham C J 1976 {\it Nucl. Phys. B} {\bf 111} 45;\hfil\break
Duff M 1977 {\it Nucl. Phys. B} {\bf 125} 334; 1984 {\it Nucl. Phys. B} {\bf 234} 
269; 1994 {\it Class. Quant. Grav.} {\bf 11} 1387.

\bibitem{tHo} `t Hooft G 1980 {\it Recent Developments in Gauge Theories}, 
$1979$ NATO Cargese advanced study institutes, Series B, Physics, v. 59,
(New York: Plenum Press);\hfil\break
See also Weinberg S 1995 {\it The Quantum Theory of Fields, Vol. II} 
(Cambridge: Cambridge University Press), Sec. 22.5, pp. 389-396.

\bibitem{DES} de Sitter W 1917 {\it Mon. Not. R. Astron. Soc} {\bf 78} 3;\hfil\break
Eddington A S 1924 {\it The Mathematical Theory of Relativity} 
(Cambridge: Cambridge University Press), pp. 161-2.

\bibitem{Sak} Sakharov A D 1966 {\it Sov. Phys. JETP} {\bf 22} 241;\\
Gliner E B 1966 {\it Sov. Phys. JETP} {\bf 22} 378.

\bibitem{Guth} Guth A H 1981, {\it Phys. Rev. D} {\bf 23}, 347.

\bibitem{WMAP} Bennett C L {\it et al.} 2003 {\it Astrophys. Jour. Supp.}, 
{\bf 148} 63;\hfil\break
Spergel D N {\it et al} 2006, e-print arxiv: astro-ph/0603449,
submitted to {\it Astrophys. J.}

\bibitem{LidL}  See {\it e.g.} Liddle A R and Lyth D H 2000, {\it Cosmological 
Inflation and Large-Scale Structure} (Cambridge: Cambridge University Press).

\bibitem{deSp} Mottola E 1985 {\it Phys. Rev. D} {\bf 31} 754.

\bibitem{deSt} Mottola E 1986 {\it Phys. Rev. D} {\bf 33} 1616.

\bibitem{Fluc} Mottola E 1986 {\it Phys. Rev. D} {\bf 33} 2136.

\bibitem{deSMM} Mazur P O and Mottola E 1986 {\it Nucl. Phys. B} {\bf 278}.

\bibitem{AMP} Anderson P R, Molina-Par\'is C and Mottola E 2005
{\it Phys. Rev. D} {\bf 72} 043515.

\bibitem{time} Mottola E 1993 {\it Physical Origins of Time Asymmetry} 
(Cambridge: Cambridge Univ. Press) Halliwell J J {\it et al.} eds., 
pp. 504-515.

\bibitem{Nob} Nobbenhuis S 2004 e-print arxiv: gr-qc/0411093, to be published in
{\it Class. Quant. Grav.}

\bibitem{Schw} Schwinger J 1951 {\it Phys. Rev.} {\bf 82} 664; 
1954 {\it Phys. Rev.} {\bf 93} 615.

\bibitem{elec} Beers B L and Nickle H H 1972 {\it Lett. Nuovo Cimento} 
{\bf 4} 320.

\bibitem{BD} Bunch T S and Davies P C W 1978 {\it Proc. R. Soc. A} {\bf 360} 117.

\bibitem{intg} Anderson J and Finkelstein D 1971 {\it Am. J. Phys.} {\bf 39} 
901;\hfil\break
van der Bij J J, van Dam H and Ng Y J 1982 {\it Physica A} {\bf 116} 307.

\bibitem{CKM} Kluger Y, Eisenberg J M, Svetitsky B, Cooper F and Mottola E 1991
{\it Phys. Rev. Lett.} {\bf 67} 2427;\hfil\break
Cooper F, Eisenberg J M, Kluger Y, Mottola E and Svetitsky B 1992 
{\it Phys. Rev.} {\bf D45} 4659.

\bibitem{brown} Einstein A 1905 {\it Ann. Physik} {\bf 17} 549.

\bibitem{LeB} Le Bellac M 1996 {\it Thermal Field Theory}
(Cambridge: Cambridge University Press).

\bibitem{GH} Gibbons G W and Hawking S W 1977 {\it Phys. Rev. D} {\bf 15} 
2738; {\it Phys. Rev. D} {\bf 15} 2752.

\bibitem{deSGP} Gibbons G W and Perry M J 1978 {\it Proc. R. Soc. London A} 
{\bf 358} 467.

\bibitem{val} Anderson P R, Molina-Par\'is C and Mottola E 2003 
{\it Phys. Rev. D} {\bf 67} 024026.

\bibitem{AIT} Antoniadis I, Iliopoulos J, and Tomaras T N 1986 {\it Phys. Rev. Lett.} 
{\bf 56} 1319.

\bibitem{AMJMP} Antoniadis I and Mottola E 1986 CERN-TH-4605/86;
1991 {\it Jour. Math. Phys.} {\bf 32} 1037.

\bibitem{AlTur} Allen B and Turyn M 1987 {\it Nucl. Phys. B} {\bf 292} 813.

\bibitem{AlFol} Allen B and Folacci A 1987 {\it Phys. Rev. D} {\bf 35} 3771.

\bibitem{Ford} Ford L 1985 {\it Phys. Rev. D} {\bf 31} 710.

\bibitem{Hig} Higuchi A and Weeks R H 2003 {\it Class. Quant. Grav.} {\bf 20} 3005.

\bibitem{MWC} Mermin N D and Wagner H 1966 {\it Phys. Rev. Lett.} {\bf 17} 1133;
\hfil\break
Coleman S 1973 {\it Commun. Math. Phys.} {\bf 31} 259.

\bibitem{Ber} Berezinskii V I 1971 {\it Sov. Phys. JETP} {\bf 32} 493.

\bibitem{FIT} Floratos E G, Iliopoulos J, and Tomaras T N 1987 {\it Phys. Lett. B}
{\bf 197} 373.

\bibitem{Wood} Tsamis N C and Woodard R P 1997 {\it Ann. Phys.} {\bf 253} 1;
1998 {\it Ann. Phys.} {\bf 267} 145;\hfil\break
Abramo L R and Woodard R P 2002 {\it Phys. Rev.} {\bf D65} 063516,
and references therein.

\bibitem{MAB} Mukhanov V, Abramo L R W and Brandenberger R 1997 {\it Phys. Rev. Lett.}
{\bf 78} 1624; \hfil\break
Abramo L R, Brandenberger R H and Mukhanov V F 1997 {\it Phys. Rev. D}
{\bf 56} 3248.

\bibitem{Unr} Unruh W 1998 e-print arxiv: astro-ph/9805069.

\bibitem{AW} Abramo L R and Woodard R P 1999 {\it Phys. Rev.  D} {\bf 60} 044010;
2002 {\it Phys. Rev. D} {\bf 65} 063516.

\bibitem{IWald} Ishibashi A and Wald R W 2006 {\it Class. Quant. Grav.} {\bf 23} 235.

\bibitem{AntM} Antoniadis I and Mottola E 1992 {\it Phys. Rev. D} {\bf 45} 
2013.

\bibitem{AMMc} Antoniadis I, Mazur P O, and Mottola E 1993 {\it Nucl. Phys. B} 
{\bf 388} 627.

\bibitem{MazMot} Mazur P O and Mottola E 2001 {\it Phys. Rev. D} {\bf 64} 104022.

\bibitem{WZ} Wess J and Zumino B 1971 {\it Phys. Lett. B} {\bf 37} 95;\hfil\break
Witten E 1979 {\it Nucl. Phys. B} {\bf 156} 269; 1983 {\it Nucl. Phys. B}
{\bf 223} 422;\hfil\break
Bardeen W A and Zumino B 1984 {\it Nucl. Phys. B} {\bf 244} 421.

\bibitem{dress} Knizhnik V G, Polyakov A M and Zamolodchikov A B 1988
{\it Mod. Phys. Lett. A} {\bf 3} 819;\hfil\break
David F 1988, {\it Mod. Phys. Lett. A} {\bf 3} 1651 (1988);\hfil\break
Distler J and Kawai H 1989 {\it Nucl. Phys. B} {\bf 321} 509.

\bibitem{DT} See {\it e.g.} Ambj{\o}rn J, Durhuus B and J\'{o}nsson T
1997 {\it Quantum Geometry: A Statistical Field Theory Approach}
(Cambridge: Cambridge University Press); \hfil\break
Ambj{\o}rn J, Carfora M and Marzuoli A 1997 {\it The Geometry of Dynamical
Triangulations} (Berlin: Springer).

\bibitem{CatMot} Catterall S and Mottola E 1999 {\it Phys. Lett. B} {\bf 467}
29; 2000 {\it Nucl. Phys. Proc. Suppl.} {\bf 83} 748.

\bibitem{Starob} Starobinsky A A 1980 {\it Phys. Lett. B} {\bf 91} 99;\hfil\break
Vilenkin A 1985 {\it Phys. Rev. D} {\bf 32} 2511;\hfil\break
Hawking S W, Hertog T and Reall H S 2001 {\it Phys. Rev. D} {\bf 63} 083504.

\bibitem{Rie} Riegert R J 1984 {\it Phys. Lett. B} {\bf 134} 56;\hfill\break
Fradkin E S and Tseytlin A A 1984 {\it Phys. Lett. B} {\bf 134} 187.

\bibitem{AMMd} Antoniadis I, Mazur P O and Mottola E 1998 {\it Phys. Lett. B} 
{\bf 444} 284.

\bibitem{AMM} Antoniadis I, Mazur P O, and Mottola E 1997 {\it Phys. Rev. D} 
{\bf 55} 4756; {\it Phys. Rev. D} {\bf 55} 4770.

\bibitem{HZ} Harrison E R 1970 {\it Phys. Rev. D} {\bf 1} 2726; \hfil\break
Zel'dovich Y B 1972 {\it Mon. Not. R. Astron. Soc.} {\bf 160} 1P.

\bibitem{sky} Antoniadis I, Mazur P O and Mottola E 1997 e-print arxiv: 
astro-ph/9705200; \hfil\break
1997 {\it Phys. Rev. Lett.} {\bf 79} 14. This paper contains a misprint in eq. (16) 
which is corrected by eq. (\ref{three}) of the present article. We thank G. Shiu 
for bringing this to our attention.

\bibitem{RG} Wilson K G and Kogut J 1974 {\it Phys. Rep. C}
{\bf 12} 75;\hfil\break
Wilson K G 1975 {\it Rev. Mod. Phys.} {\bf 47} 773.

\bibitem{pol} Polyakov A M 1970 {\it Sov. Phys. JETP Lett.} {\bf 12} 381.

\bibitem{peeb} Peebles P J E 1993 {\it Principles of Physical Cosmology} 
(Princeton: Princeton University Press).

\bibitem{spike} Antoniadis I, Mazur P O and Mottola E 1997
{\it Phys. Lett. B} {\bf 394} 49.

\bibitem{Kom} Komatsu E {\it et al.} 2003 {\it Astrophys. J. Suppl.}
{\bf 148} 119.

\bibitem{DGH} Donoghue J F, Golowich E and Holstein B R 1992
{\it Dynamics of the Standard Model} (Cambridge: Cambridge University Press)
Chap. IV, Secs. VI-5 and VII-3, and references therein.

\bibitem{Dono} Donoghue J F 1994 {\it Phys. Rev. D} {\bf 50} 3874.

\bibitem{Adler} Adler S L 1969 {\it Phys. Rev.} {\bf 177} 2426;\hfil\break
Bell J S and Jackiw R 1969 {\it Nuovo Cimento A} {\bf 60} 47.

\bibitem{BGMF} Bardeen W A, Fritzsch H and Gell-Mann M 1973 in
{\it Scale and Conformal Symmetry in Hadron Physics} (New York: Wiley),
Gatto R ed. 

\bibitem{Wein} Cheng T P and Li L F 1984 {\it Gauge Theory of Elementary Particle
Physics} (Oxford: Oxford University Press) Sec. 6.2, pp. 173-182;
\hfil\break
Weinberg S 1995 {\it The Quantum Theory of Fields, Vol. II} (Cambridge:
Cambridge University Press) Sec. 22.2, pp. 362-370.

\bibitem{Odint} Odintsov S D 1992 {\it Z. Phys. C} {\bf 54} 531;\hfil\break
Antoniadis I and Odintsov S D 1995 {\it Phys. Lett. B} {\bf 343} 76.

\bibitem{BFS}  Balbinot R, Fabbri A and Shapiro I L 1999 {\it Phys. Rev. Lett.}
{\bf 83} 1494; 1999 {\it Nucl. Phys. B} {\bf 559} 301.

\bibitem{MotV} Mottola E and Vaulin R 2006 {\it Phys. Rev. D} {\bf 74} 064004.

\bibitem{HarHaw} Hartle J B and Hawking S W 1976 {\it Phys. Rev. D} {\bf 13}
2188;\hfil\break
Israel W 1976 {\it Phys. Lett. A} {\bf 57} 107.

\bibitem{Boul} Boulware D G 1975 {\it Phys. Rev. D} {\bf 11} 1404;
1976 {\it Phys. Rev. D} {\bf 13} 2169.

\bibitem{ChrFul} Christensen S M and Fulling S A 1977 {\it Phys. Rev. D}
{\bf 15} 2088.

\bibitem{CanDeu} Deutsch D and Candelas P 1979 {\it Phys. Rev. D} {\bf 20} 3063.

\bibitem{CHLS} Chapline G, Hohlfield E, Laughlin R B and Santiago D I 2001 
{\it Phil. Mag. B} {\bf 81} 235.

\bibitem{star} Mazur P O and Mottola E 2001 e-print arxiv: gr-qc/0109035;\hfil\break
2004 {\it Proc. Nat. Acad. Sci.} {\bf 101} 9545.

\bibitem{Boy} Boyer T H 1968 {\it Phys. Rev.} {\bf 174} 1764.

\bibitem{Dir} Dirac P A M 1962 {\it Proc. Roy. Soc. Lond.} {\bf 270} 354.

\bibitem{Vis} Visser M and Wiltshire D L 2004 {\it Class. Quant. Grav.} {\bf 21} 1135;
\hfil\break
Cattoen C, Faber T and Visser M 2005 {\it Class. Quant. Grav.} {\bf 22} 4189.

\bibitem{Maz} Mazur P O 1996 {\it Acta Phys. Polon. B} {\bf 27} 1849;
1997 e-print arxiv: hep-th/9712208.

\end{thebibliography}
\end{document}